\documentclass[10pt,journal,compsoc]{IEEEtran}
\ifCLASSOPTIONcompsoc
  \usepackage[nocompress]{cite}
\else
  \usepackage{cite}
\fi
\ifCLASSINFOpdf
  \usepackage[pdftex]{graphicx}
  \graphicspath{{fig/}{./}} % where to search for the images
  \DeclareGraphicsExtensions{.pdf,.jpeg,.jpg,.png}
\else
  \usepackage[dvips]{graphicx}
  \graphicspath{{../eps/}}
  \DeclareGraphicsExtensions{.eps}
\fi
\usepackage{amsmath}
\usepackage{tabularx}

\usepackage[compatibility=false]{caption}
\usepackage{subcaption}

\usepackage{url}
\usepackage{hyperref}
\usepackage{microtype}                 % use micro-typography (slightly more compact, better to read)
\PassOptionsToPackage{warn}{textcomp}  % to address font issues with \textrightarrow
\usepackage{textcomp}                  % use better special symbols
\usepackage{mathptmx}                  % use matching math font
\usepackage{times}                     % we use Times as the main font
\usepackage{tabu}                      % only used for the table example
\usepackage{booktabs}                  % only used for the table example
\usepackage{color, soul}
\usepackage[percent]{overpic}
\usepackage{flushend}
\usepackage{enumitem}% http://ctan.org/pkg/enumitem
\usepackage{comment}
\usepackage{xcolor}
\setlist{nosep}
\hypersetup{%
    pdfborder = {0 0 0},
    pdfauthor={Ga\"elle Richer, 
      Alexis Pister, 
      Moataz Abdelaal, 
      Jean-Daniel Fekete, 
      Michael Sedlmair,
      and Daniel Weiskopf},
    pdftitle={Scalability in Visualization},
    pdfkeywords={scalability, visualization, structured literature analysis, conceptual framework}
}

\usepackage{tikz}
\usepackage{tcolorbox}
\usetikzlibrary{positioning}
\tikzstyle{edge}=[black,->,thick,line width=0.25mm] 
\tikzstyle{edgetext}=[minimum width=3cm, text width=2.8cm,font=\sffamily\scriptsize\itshape] 
\tikzstyle{mainbox}=[font=\large,draw=black,minimum width=1cm,minimum height=1cm,align=center] 

\newcommand{\modelbox}{
    \node[mainbox]  (box) at (0, 0) {$f$};
}

\newcommand{\modeloneps}[1]{
    \draw[edge,<-]  (box.180) -- ++(-0.3,0) node [edgetext,left,align=right] {#1};
}

\newcommand{\modeltwops}[2]{
    \draw[edge,<-]  (box.160) -- ++(-0.3,0)  node [edgetext,left,align=right] {#1};
    \draw[edge,<-]  (box.-160) -- ++(-0.3,0) node [edgetext,left,align=right] {#2};
}

\newcommand{\modeloneef}[1]{
    \draw[edge,->]  (box.0) -- ++(0.3,0) node [edgetext,right,align=left,anchor=west] {#1};
}

\newcommand{\modeltwoef}[2]{
    \draw[edge,->] (box.20) -- ++(0.3,0)  node [edgetext,right,align= left,anchor=west] {#1};
    \draw[edge,->] (box.-20) -- ++(0.3,0) node [edgetext,right,align=left,anchor=west] {#2};
}

\newcommand{\modeloners}[1]{
    \draw[edge,<-] (box.north) -- (0,0.8) node[edgetext, align=center] at (0,1) {#1};
}

\newenvironment{cartouche}[3]
    {\begin{tcolorbox}[size=title,boxrule=0.2pt]
    \sffamily\scriptsize
    \textbf{Example}: #1
    \tcblower
    \sffamily\scriptsize
    \textit{Expression}: #2\\
    \textit{Meaning}: #3
    \begin{center}
    \begin{tikzpicture}[font=\sffamily\scriptsize]
    }
    {\end{tikzpicture} \end{center}\end{tcolorbox}}

\newcommand{\nfiltered}{157} % Previously 146
\newcommand{\nexcluded}{30}
\newcommand{\nbarely}{14}
\newcommand{\ncoded}{127}  % coded=filtered-excluded
\newcommand{\nconsensus}{21}
\newcommand{\nredundant}{106} %consensus+redundant=coded
\newcommand{\totvispub}{3394}

\newcommand{\etal}{\textit{et al}.} % Use et al. when three or more names are given for a reference cited in the text. 
\newcommand{\ie}{\textit{i}.\textit{e}.,\ }
\newcommand{\eg}{\textit{e}.\textit{g}.,\ }

\newcommand{\say}[1]{\itshape ``#1''}
\definecolor{verylightgray}{RGB}{235,235,235}
\newcommand{\code}[1]{\scalebox{.9}[1]{\footnotesize\sffamily{\sethlcolor{verylightgray}\hl{#1}}}}

\usepackage[normalem]{ulem}
\newcommand{\add}[1]{#1}
\newcommand{\del}[1]{}

\ifdefstring{\mode}{allchanges}{
    \renewcommand{\add}[1]{\textcolor{green!50!black}{#1}}
    \renewcommand{\del}[1]{\textcolor{black!50!red}{\sout{#1}}}
    
}
\ifdefstring{\mode}{additions}{
    \renewcommand{\add}[1]{\textcolor{green!50!black}{#1}}
}

\begin{document}

% paper title
% Titles are generally capitalized except for words such as a, an, and, as,
% at, but, by, for, in, nor, of, on, or, the, to and up, which are usually
% not capitalized unless they are the first or last word of the title.
% Linebreaks \\ can be used within to get better formatting as desired.
% Do not put math or special symbols in the title.
\title{Scalability in Visualization}
%
%
% author names and IEEE memberships
% note positions of commas and nonbreaking spaces ( ~ ) LaTeX will not break
% a structure at a ~ so this keeps an author's name from being broken across
% two lines.
% use \thanks{} to gain access to the first footnote area
% a separate \thanks must be used for each paragraph as LaTeX2e's \thanks
% was not built to handle multiple paragraphs
%
%
%\IEEEcompsocitemizethanks is a special \thanks that produces the bulleted
% lists the Computer Society journals use for "first footnote" author
% affiliations. Use \IEEEcompsocthanksitem which works much like \item
% for each affiliation group. When not in compsoc mode,
% \IEEEcompsocitemizethanks becomes like \thanks and
% \IEEEcompsocthanksitem becomes a line break with idention. This
% facilitates dual compilation, although admittedly the differences in the
% desired content of \author between the different types of papers makes a
% one-size-fits-all approach a daunting prospect. For instance, compsoc 
% journal papers have the author affiliations above the "Manuscript
% received ..."  text while in non-compsoc journals this is reversed. Sigh.

\author{Ga\"elle Richer,
  Alexis Pister, 
  Moataz Abdelaal, 
  Jean-Daniel Fekete,  
  Michael Sedlmair,
  and Daniel Weiskopf%

  \IEEEcompsocitemizethanks{\IEEEcompsocthanksitem G. Richer, A. Pister, and J.-D. Fekete are with Universit\'e Paris-Saclay, CNRS, Inria, LISN, France. A. Pister is also with I3, CNRS, Telecom Paris, Institut Polytechnique de Paris, France.\protect\\
% note need leading \protect in front of \\ to get a newline within \thanks as
% \\ is fragile and will error, could use \hfil\break instead.
E-mail: \{first-name.last-name\}@inria.fr
\IEEEcompsocthanksitem M. Adbelaal, M. Sedlmair, and D. Weiskopf are with University of Stuttgart, Germany.\protect\\
E-mail: \{first-name.last-name\}@visus.uni-stuttgart.de}% <-this % stops an unwanted space
\thanks{Manuscript received xx xxx. 201x; accepted xx xxx. 201x. Date of Publication
xx xxx. 201x; date of current version xx xxx. 201x. For information on
obtaining reprints of this article, please send e-mail to: reprints@ieee.org.
Digital Object Identifier: xx.xxxx/TVCG.201x.xxxxxxx.}
}

% note the % following the last \IEEEmembership and also \thanks - 
% these prevent an unwanted space from occurring between the last author name
% and the end of the author line. i.e., if you had this:
% 
% \author{....lastname \thanks{...} \thanks{...} }
%                     ^------------^------------^----Do not want these spaces!
%
% a space would be appended to the last name and could cause every name on that
% line to be shifted left slightly. This is one of those "LaTeX things". For
% instance, "\textbf{A} \textbf{B}" will typeset as "A B" not "AB". To get
% "AB" then you have to do: "\textbf{A}\textbf{B}"
% \thanks is no different in this regard, so shield the last } of each \thanks
% that ends a line with a % and do not let a space in before the next \thanks.
% Spaces after \IEEEmembership other than the last one are OK (and needed) as
% you are supposed to have spaces between the names. For what it is worth,
% this is a minor point as most people would not even notice if the said evil
% space somehow managed to creep in.

% The paper headers
\markboth{Scalability in Visualization}%
{Richer \MakeLowercase{\etal}: Scalability in Visualization}
% The only time the second header will appear is for the odd numbered pages
% after the title page when using the twoside option.
% 
% *** Note that you probably will NOT want to include the author's ***
% *** name in the headers of peer review papers.                   ***
% You can use \ifCLASSOPTIONpeerreview for conditional compilation here if
% you desire.

% The publisher's ID mark at the bottom of the page is less important with
% Computer Society journal papers as those publications place the marks
% outside of the main text columns and, therefore, unlike regular IEEE
% journals, the available text space is not reduced by their presence.
% If you want to put a publisher's ID mark on the page you can do it like
% this:
%\IEEEpubid{0000--0000/00\$00.00~\copyright~2015 IEEE}
% or like this to get the Computer Society new two part style.
%\IEEEpubid{\makebox[\columnwidth]{\hfill 0000--0000/00/\$00.00~\copyright~2015 IEEE}%
%\hspace{\columnsep}\makebox[\columnwidth]{Published by the IEEE Computer Society\hfill}}
% Remember, if you use this you must call \IEEEpubidadjcol in the second
% column for its text to clear the IEEEpubid mark (Computer Society jorunal
% papers don't need this extra clearance.)

% use for special paper notices
%\IEEEspecialpapernotice{(Invited Paper)}

% for Computer Society papers, we must declare the abstract and index terms
% PRIOR to the title within the \IEEEtitleabstractindextext IEEEtran
% command as these need to go into the title area created by \maketitle.
% As a general rule, do not put math, special symbols or citations
% in the abstract or keywords.
\IEEEtitleabstractindextext{%
\begin{abstract}
We introduce a conceptual model for scalability designed for visualization research. With this model, we systematically analyze over 120 visualization publications from 1990 to 2020 to characterize the different notions of scalability in these works. 
While many papers have addressed scalability issues, our survey identifies a lack of consistency in the use of the term in the visualization research community.
We address this issue by introducing a consistent terminology meant to help visualization researchers better characterize the scalability aspects in their research.
It also helps in providing multiple methods for supporting the claim that a work is ``scalable.''
Our model is centered around an effort function with inputs and outputs.
The inputs are the problem size and resources, whereas the outputs are the actual efforts, for instance, in terms of computational run time or visual clutter.
We select representative examples to illustrate different approaches and facets of what scalability can mean in visualization literature. 
Finally, targeting the diverse crowd of visualization researchers without a scalability tradition, we provide a set of recommendations for how scalability can be presented in a clear and consistent way to improve fair comparison between visualization techniques and systems and foster reproducibility.
\end{abstract}

% Note that keywords are not normally used for peerreview papers.
}

% make the title area
\maketitle

% To allow for easy dual compilation without having to reenter the
% abstract/keywords data, the \IEEEtitleabstractindextext text will
% not be used in maketitle, but will appear (i.e., to be "transported")
% here as \IEEEdisplaynontitleabstractindextext when the compsoc 
% or transmag modes are not selected <OR> if conference mode is selected 
% - because all conference papers position the abstract like regular
% papers do.
\IEEEdisplaynontitleabstractindextext
% \IEEEdisplaynontitleabstractindextext has no effect when using
% compsoc or transmag under a non-conference mode.

% For peer review papers, you can put extra information on the cover
% page as needed:
% \ifCLASSOPTIONpeerreview
% \begin{center} \bfseries EDICS Category: 3-BBND \end{center}
% \fi
%
% For peerreview papers, this IEEEtran command inserts a page break and
% creates the second title. It will be ignored for other modes.
\IEEEpeerreviewmaketitle

\IEEEraisesectionheading{\section{Introduction}\label{sec:introduction}}
% Computer Society journal (but not conference!) papers do something unusual
% with the very first section heading (almost always called "Introduction").
% They place it ABOVE the main text! IEEEtran.cls does not automatically do
% this for you, but you can achieve this effect with the provided
% \IEEEraisesectionheading{} command. Note the need to keep any \label that
% is to refer to the section immediately after \section in the above as
% \IEEEraisesectionheading puts \section within a raised box.

% The very first letter is a 2 line initial drop letter followed
% by the rest of the first word in caps (small caps for compsoc).
% 
% form to use if the first word consists of a single letter:
% \IEEEPARstart{A}{demo} file is ....
% 
We address the issue of characterizing \emph{scalability} in visualization research. Scalability is a frequent topic, with many papers claiming to improve scalability or achieve scalable---or sometimes, more scalable---techniques. The visualization research community has a long tradition of acknowledging the need for scalable solutions, as for example, included in summaries of grand research challenges for various communities of visualization~\cite{DBLP:journals/cga/Johnson04,DBLP:journals/cga/WongSJCR12} or roadmaps for future research~\cite{Thomas:2005:illumpath,DBLP:books/daglib/0028506}. 

Despite the high relevance of scalability---or maybe, because of this---we noticed a large range of connotations or uses of this concept in visualization papers. This situation reflects the large diversity of research topics and methods in visualization, and it may also be the multidisciplinary nature of visualization, which includes research from computer science and algorithms, human-computer interaction, psychology, etc. Some of these communities have established models and methods for assessing scalability, but not all of them. However, even when such approaches might be established in another community, they might not necessarily be common knowledge in the visualization research community. Furthermore, it is not always clear whether a wholesale adoption of these methods is possible or if they need to be adapted and fine-tuned to the specificities of visualization research. 
 
In short, there is a wide range of interpretations of the concept of scalability in visualization, sometimes with only implicit documentation and communication of the concrete interpretation used in a paper. This can lead to misunderstandings and impair the reproducibility of research results.

The recent restructuring of the IEEE VIS conferences into a single conference with multiple areas attests that visualization research is becoming more diverse and trying to be more integrated. While some articles will remain targeted to a distinct audience well aware of its own meaning of scalability, a growing number of articles will cross boundaries to address multiple meanings of scalability, leading to more diverse reviewers and readers, with different backgrounds. We aim at helping authors, reviewers, and readers navigate the different aspects of scalability.

To this end, we contribute a \emph{conceptual model for scalability} that is designed to be versatile and flexible enough to capture existing uses of the concept `scalability' in visualization research, align terminology, improve conceptual and methodological consistency across domains, and allow for other uses in the future. 
\add{In particular, we envision the model to help  communicate about scalability across the diverse subcommunities of visualization. }
Our model is built on an \emph{effort} function that takes inputs in the form of problem size, assumptions, and descriptions of resources, and maps these to a description of effort as the output associated with the visualization. Key to the flexibility of the model is the large freedom in modeling the inputs and outputs: they can cover technical aspects such as data set size, available compute nodes, or compute times, all the way to human-oriented aspects like readability or user task performance. Therefore, we are able to show that this model can be instantiated to cover the typical scenarios of scalability in visualization, and also the different interpretations of the terms ``scalability,'' ``scalable,'' and ``more scalable.'' 

\add{We argue that seeking the common traits between the multiple existing definitions and presenting them with a unified model creates a helpful framework for comprehension. We recognize our model would not help authors and reviewers within their subfield, \eg visualization in high-performance computing (HPC) or graph drawing, since they have a clear understanding of their own meaning of scalability. However, it becomes useful for articles mixing two aspects of scalability, \eg how HPC can provide more readable features, with a need to be understood by the two subcommunities. 
%We argue that 
This kind of scenario is becoming more frequent in visualization and this is why we need a unifying model.}

\del{We ground our conceptual model in the current state of visualization research. To this end, we} \add{Using the conceptual model as a framework, we analyzed the current state of visualization research and} contribute a \emph{structured and systematic literature analysis} of the full papers published in IEEE Visualization, SciVis, InfoVis, and VAST from 1990 to 2020. The literature search led to \ncoded\ articles for which we derived a coding scheme and analyzed them, respectively. Four of the authors participated in multiple rounds of reviews of these relevant papers followed by discussions to establish the conceptual model, scenarios, and coding scheme. The two other authors coded the complete set of papers after being introduced to the coding scheme. Our goal was to learn about the current usage of the notion of scalability in visualization research, as well as to assess how well our conceptual model allows characterizing previous research on scalability. 
We make the coding book and results publicly available \add{at the following repository: \texttt{\url{https://osf.io/xrvu7/}}}.

Based on our conceptual model, general observations, and the literature review, we arrive at {\em recommendations} to improve the design and presentation of scalability-related research \add{when targeting an outside or mixed audience}. We believe that this would \add{also help} compare visualization techniques and systems, and foster reproducibility.

\section{Related Work}\label{sec:related-work}

The visualization research community has become more diverse over the years, starting with statistics, algorithms, computer graphics, and computational science in the early 1990s, and joined by human-computer interaction (HCI), psychology, vision science, design, cartography, and many more. The concept of scalability varies from one community to the next, with different levels of maturity.
In this section, we review related work discussing and defining scalability in different areas of computer science and in the visualization community.

\subsection{Definitions of Scalability}

Weinstock and Goodenough~\cite{weinstock2006system} define the scalability problem as ``the inability of a system to accommodate an increased workload.''
Bondi~\cite{DBLP:conf/wosp/Bondi00} mentions several definitions of scalability in computer science:
\begin{itemize}[nosep]
    \item ``\emph{Scalability} is the property of a system to handle a growing amount of work by adding resources to the system.'' Adding resources may have the form of adding more nodes to a system made of multiple small interconnected servers (scaling \emph{out} or \emph{horizontally}) or adding more resources to a single node (scaling \emph{up} or \emph{vertically})~\cite{DBLP:conf/ipps/MichaelMSW07}.
    \item \emph{Load scalability} is the ``ability to function gracefully, \ie without undue delay and without unproductive resource consumption or resource contention at light, moderate, or heavy loads while making good use of available resources.''
    \item \emph{Space scalability} is that ``memory requirements do not grow to intolerable levels as the number of items it supports increase.''
    \item \emph{Space-time scalability}: ``continues to function gracefully as the number of objects [\ldots] increases by orders of magnitude.''
    \item \emph{Structural scalability} means that ``implementation or standards do not impede the growth of the number of objects it encompasses, or at least will not do so within a chosen time frame.''
\end{itemize}
\smallskip
    
\noindent Parallel systems and HPC distinguish mainly two types of scalability:
\begin{itemize}
    \item \emph{Strong scaling}: ``how the solution time varies with the number of processors for a fixed total problem size.''
    \item \emph{Weak scaling}: ``how the solution time varies with the number of processors for a fixed problem size per processor.''
\end{itemize}
\smallskip
    
\noindent Hill~\cite{hill1990scalability} tries to define scalability for multiprocessor systems and admits: ``but I fail to find a useful, rigorous definition of it.''
\noindent Duboc \etal~\cite{duboc2006scalability} define it as: ``a quality of software systems characterized by the causal impact that scaling aspects of the system environment and design have on certain measured system qualities as these aspects are varied over expected operational ranges. If the system can accommodate this variation in a way that is acceptable to the stakeholder, then it is a scalable system.'' 

All the definitions are specified as properties of systems at an abstract level, focusing on ``amount of work,'' ``delay,'' ``resources,'' ``productive resource consumption,'' ``[work]loads,'' ``memory,'' ``function gracefully,'' ``time frame,'' ``adding nodes,'' and ``shared memory.''
They rely on implicit domain knowledge to be clearly understood and are not suitable to the wide audience of visualization practitioners. 

\subsection{Scalability in Visualization and Visual Analytics}

Visualization and visual analytics are concerned with general computer science scalability when it comes to systems or algorithms. 
In addition, they are also concerned with more specific issues. Robertson \etal~\cite{DBLP:journals/ivs/RobertsonEEKJ09} mention information scalability, visual scalability, display scalability, and human scalability, in addition to computational scalability. They also add other scalability issues: software scalability, temporal scalability, cross-scale issues, privacy and security issues (related to scale), and language issues.
Yost and North~\cite{DBLP:journals/tvcg/YostN06} also mention graphical scalability (``limits imposed by the number of pixels'') and perceptual scalability (``When the screen is not the limiting factor, just how much data can a person effectively perceive?'').
Eick and Karr~\cite{eick2002visual} want to quantify visual scalability by modeling the dependence between responses, factors, and data.
They admit that it cannot be done because few responses can be quantified or measured. Instead, they break down the problem into subparts affecting the overall scalability, adding ``visual metaphors,'' ``interactivity,'' and ``aggregation'' to the list of factors affecting scalability.

Scalability is also related to evaluation since it is based on measuring efficiency. Lam \etal~\cite{DBLP:journals/tvcg/LamBIPC12} describe seven scenarios for evaluation in visualization, some of them leading to quantitative results and others to qualitative ones. Scalability is part of the ``Evaluating User Performance" and ``Evaluating Visualization Algorithms" scenarios. 
One area in which scalability evaluation is well-established is the HPC/visualization community, where the main focus is on algorithmic scalability with well-defined metrics and definitions (\eg strong scalability). However, the rest of the visualization community may not be familiar with these definitions, and it remains unclear if they could be applied in a broader context than those with HPC resources.

\subsection{Scalability in HCI, Psychology, and Vision Science}

Scalability related to humans is different from scalability in computer science.
In their seminal book, Card \etal~\cite{DBLP:books/lib/CardMN83} describe the human as a processor with numerous capabilities, some of them ruled by laws or models expressible mathematically.
Visualization is concerned with several of these capabilities, in particular regarding perceptual scalability, cognitive scalability, and movement. The psychology laws and models often refer to information theory, considering perception and action as communication through capacity-limited channels.

Scalability has been studied for some aspects of visual perception, such as ensemble coding~\cite{Szafir2016ensemble}, preattentive processing~\cite{Treisman85preattentive}, and the limit of the number of colors perceived efficiently~\cite{healey1999},
Fitts' law~\cite{Fitts1954} for pointing, the scalability of item selection and navigation~\cite{DBLP:conf/bcshci/GuiardBMB01},  
Hick's Law~\cite{HickLaw} for reading items, and the scalability of menus~\cite{Cockburn2007menu}.
Budiu~\cite{Budiu2014scalingtheui} discusses several issues related to scaling user interfaces: working memory limits, screen size that limits the capacity of the communication channel, and attention limits. 
Brown~\etal~\cite{Brown2017HCIScale} list scalability challenges in HCI relative to the number of users, the different contexts of use, and the multiplicity of systems and technologies.
Therefore, while most of the human capabilities exhibit hard limits, interaction and visualization techniques allow performing tasks with various interpretations of scalability. 

\subsection{Summary}

Scalability is addressed in many ways by the different disciplines and communities related to visualization. Still, they share many concepts but instantiate these concepts with wide variations.

Several articles relate scalability to ``factors and certain dependent variables''~\cite{duboc2006scalability,eick2002visual,ruthruff2006}, also called independent variables and measures. Duboc \etal~\cite{duboc2006scalability} also mention \emph{nuisance variables}: ``Variables whose effects cannot be completely controlled for or variables that are simply not considered in the experiment design.'' They also consider the scalability problem as a multi-criteria optimization problem with multiple measures to optimize, combined into a utility function. Although we acknowledge that their model is useful, we believe it is too complicated with respect to the interpretations of scalability as seen in the visualization research community where the measures are not usually combined. 
% PAGE LIMIT EMERGENCY BUTTON \jdf{Feel free to remove the following lines.}
\begin{comment}

This article tackles the challenge of Hill~\cite{hill1990scalability} in a slightly different direction: instead of defining scalability, we provide a model to characterize it.

\begin{quotation}
``I examined aspects of scalability, but did not find
a useful, rigorous definition of it. Without such a
definition, I assert that calling a system ‘scalable’
is about as useful as calling it ‘modern’. I encourage the technical community to either rigorously
define scalability or stop using it to describe systems.''
\end{quotation}
---Mark D. Hill, What is Scalability?~\cite{hill1990scalability}

\end{comment}

Human capabilities do not scale as nicely as machine ones. Visual perception is limited in scalability by physiological factors, such as the number of cones and rods at the lower level. Some pattern processing allows humans to perform important tasks efficiently (sometimes called ``preattentively''), but these perceptual tasks only work under stringent limits. The eye can track the movement of a few moving objects on a screen, but this tracking fails when too many objects cross (visual crowding). Therefore, scalability-related human performance can hardly be assessed on theoretical grounds only, it should usually be checked with experiments. For these reasons, while scalability in the HPC and distributed computing communities has established definitions and evaluation methodologies, these do not directly translate to all parts of a visualization system with a human in the loop.

In this article, we do not define scalability but provide a model to express particular instances of scalability according to the ``utility function'' of Duboc \etal~\cite{duboc2006scalability}.

\section{Scalability Model}\label{sec:model}

Our first contribution is a conceptual model that describes the scalability of a visualization system, component, or technique. 
The model is designed to (1) express different scalability concerns that are relevant to visualization applications (\eg visual, perceptual, computational), (2) be applied to different parts of the visualization pipeline, and (3)  allow reasoning about different meanings of \textit{scalable} and \textit{scalability}.

\subsection{Model Components}
\begin{figure}[b]
    \centering
    \begin{tikzpicture}[font=\sffamily\small,thick]
        \modelbox
        \modeloneps{problem sizes $S$}
        \modeloneef{efforts $E$}
        \modeloners{resources $R$}
        \draw[edge,dashed,<-] (0,-0.5) -- ++(0,-0.4) node[font=\sffamily\scriptsize\itshape] at (0,-1.1) {assumptions $A$};
    \end{tikzpicture}
\caption{Conceptual model with problem size variables $S$ and resource variables $R$ as input, assumptions $A$, and effort variables $E$ as output to $f$.
}\label{fig:model-diagram}
\end{figure}
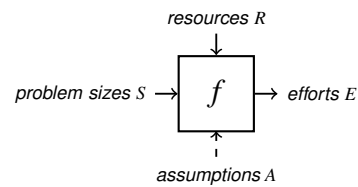

The scalability model represents the scalability of a \textit{visualization process} that tackles a specific problem, by a function with four components: \emph{problem size}, \emph{resources}, \emph{assumptions}, and \emph{effort}, which are described in more detail in the coming subsections.
The function maps the problem size, expected to vary or grow across applications, to the effort associated with the process's solution to the problem, provided an amount of resources and some assumptions, specific to the particular problem addressed. 
The relationship between these four components is formalized by the function $f$:
\[
    f:(S; R, A) \longmapsto E
\]
with $S$ being the set of variables describing the problem size, $R$ describing the available resources, $A$ the assumptions, and $E$ the effort associated with the result. The components of the conceptual model are summarized in ~\autoref{fig:model-diagram}. The notation separates $S$ from $R$ and $A$ to express the difference in role of the actual input $S$ from the context parameters $R$ and $A$.

\subsubsection{Problem Sizes \textit{S}}
The \emph{problem size} variables are properties that characterize the complexity of the problem targeted or solved by the process. Most commonly, these variables are descriptions of the size of the input data: either in number of elements or attributes for discrete data, or in sample size for continuous data. However, they could also correspond to data characteristics that go beyond data size, such as data distribution, or refer to input other than data, such as the number of simultaneous users or the visual output size (\eg image resolution).

\subsubsection{Resources \textit{R}}
The \emph{resource} variables are properties related to the material components of the system or application environment. They are factors influencing effort while being independent of the input data. They typically include computational resources (\eg number of cores, memory), %number of pixels on the display, 
or other resources that the designer can leverage to improve performance. Resources are characterized by the fact that they are often limited in practice, and therefore the optimization of their usage is one lever to improve performance.

In some communities like HPC, being scalable encompasses the intent to optimize resource usage as well as being designed to gracefully adapt and make use of any additional resource available at their maximum capacity. Examples include networks of computers (\eg \cite{perrot2015large}) or grids of projectors (\eg \cite{chen2002scalable}). In other communities like HCI, having additional screens is related to opportunity since screens are relatively cheap and can also be shared between applications. Scalability questions relate to the usefulness of dedicating more screens to a visualization application when the screens are already available (\eg \cite{jakobsen2011}). 
Depending on the community, using multiple processor cores is considered as resource optimization or opportunity. We connect the resources and meaning of scalability in more detail in \autoref{sec:meaning}. 

\subsubsection{Assumptions \textit{A}}
\emph{Assumptions} define the validity bounds of the function $f$ for the chosen research context and problem definition. More precisely, assumptions include the range of available resources, the range of reasonable values to expect for problem size, and the range of values considered acceptable or satisfactory for effort variables (\eg interaction rates perceived as interactive~\cite{DBLP:books/lib/CardMN83}). 

\subsubsection{Efforts \textit{E}}
The \emph{effort} variables are properties describing the performance of the process. Effort variables can be measures of efficiency (\eg readability) or measures of cost (\eg computation time). For convenience, we consider that effort variables can always be expressed as a cost, where lower values are better. 
From a computational perspective, some examples of effort variables are computation time or frame rate. From a visual and user perspective, some examples are the ambiguity of a data representation (as opposed to its faithfulness), the interactivity of the system, the ease and speed of task completion, the number of insights, or visualization quality metrics. 
As in complexity theory, across all applications, effort can be defined as average, best, or worst case for a given set of problem size characteristics. Additionally, other aggregates across outcomes could be considered, such as standard deviation, for instance. 

\subsection{Meaning and Expression}\label{sec:meaning}
\begin{figure}
    \centering
     \begin{subfigure}[t]{0.32\columnwidth}
        \includegraphics[width=.75\textwidth]{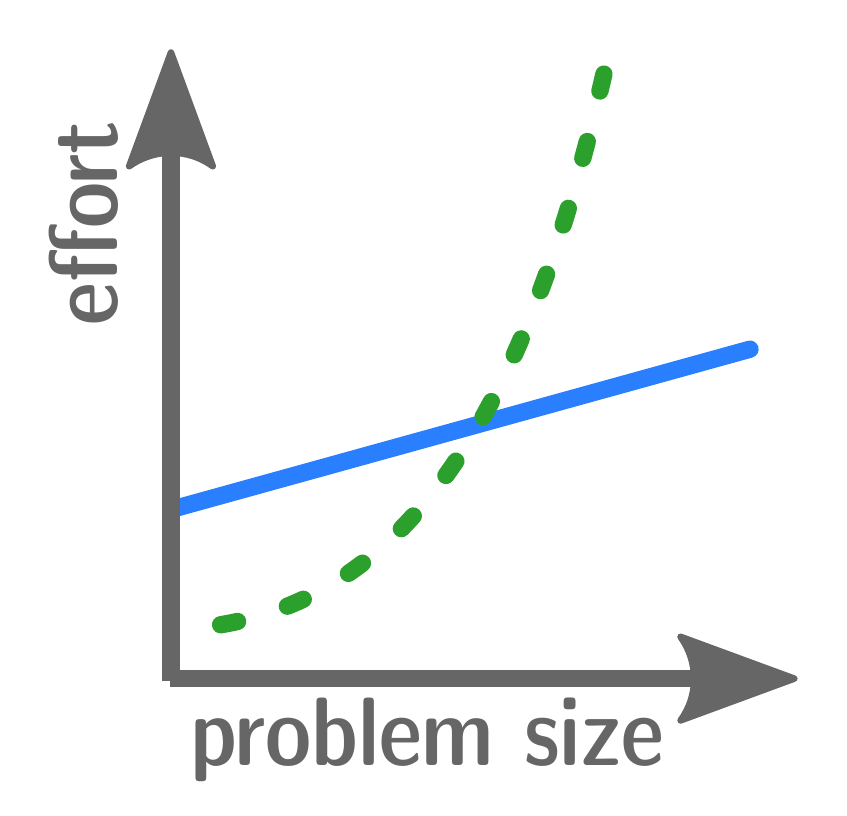}
        \caption{Shape.}
        \label{fig:meaning-shape}
     \end{subfigure}\hfill
     \begin{subfigure}[t]{0.32\columnwidth}
        \includegraphics[width=.75\textwidth]{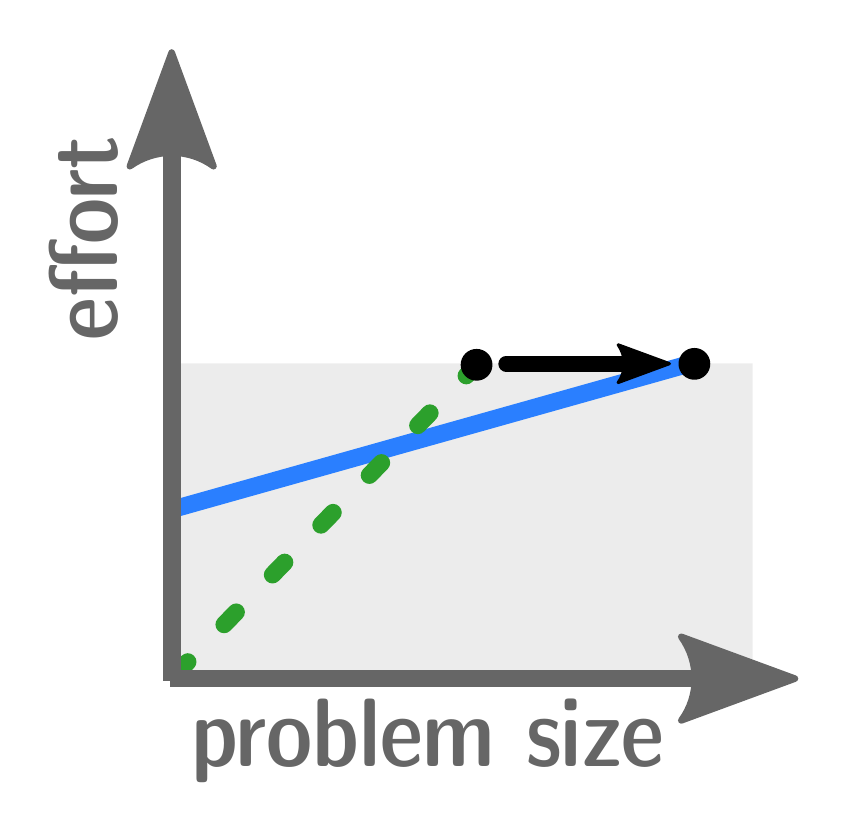}
        \caption{Threshold.}
        \label{fig:meaning-threshold}
     \end{subfigure}\hfill
      \begin{subfigure}[t]{0.33\columnwidth}
        \includegraphics[width=.75\textwidth]{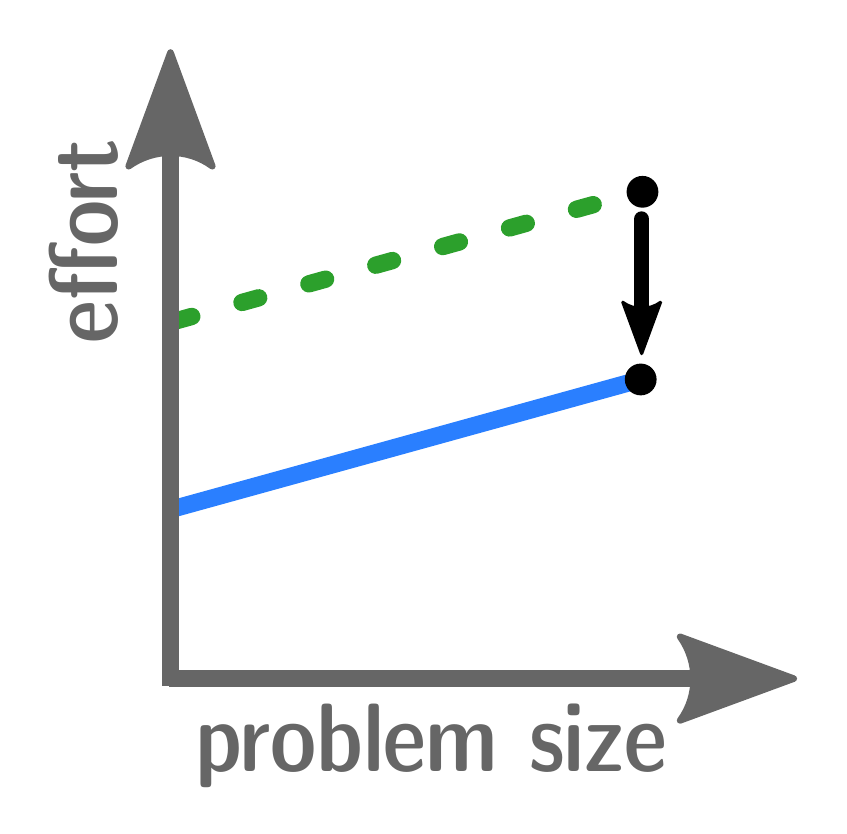}
        \caption{Better performance.}
        \label{fig:meaning-better-performance}
     \end{subfigure}
     \caption{Examples of effort functions, $f_{\text{new}}$ in solid blue and $f_{\text{old}}$ in dashed green, with  $f_{\text{new}}$ being more scalable than $f_{\text{old}}$ according to three meanings of scalable.}
     \label{fig:some-meanings}
\end{figure}

Generally, a scalability issue is an inability of a technique or system to accommodate an increase in problem size, for the given resources. The inability is manifested by efforts that do not meet requirements, \eg inaccurate results, processing that takes too long, or a system failing to respond. A \textit{scalable} system might address a scalability issue in different ways. 
In our model, the function $f$ formalizes the relationship between the components relevant to scalability such that the \textit{meaning} of scalability can be expressed as properties of $f$. We present three examples of meanings of \textit{scalable} or \textit{more scalable} in relation to $f$, illustrated in~\autoref{fig:some-meanings}: based on the shape of $f$, based on $f$ passing a threshold, or on $f$ demonstrating better performance on large problem sizes. 

When scalability is the ability to sustainably handle increasingly large problem sizes with reasonable effort, it corresponds to concluding scalability from the \textit{shape} of $f$. One example, illustrated in~\autoref{fig:meaning-shape}, is demonstrating that $f$ (blue line) is linear with respect to increasing problem size at fixed resources. 
Another example is demonstrating that $f$ is reciprocal with respect to increasing amount of resources at a fixed problem size, to show that an increase in problem size can be accommodated by more resources.

When scalability is the ability to handle problems of larger size than before, it corresponds to the function $f_{\text{new}}$ sustaining the same amount of effort for a larger problem size than another $f_{\text{old}}$:
\[
    f_{\text{new}}(S_{\text{new}}; R, A) = f_{\text{old}}(S_{\text{old}}; R, A) = E_{\text{fixed}}
\]
with $S_{\text{new}}>S_{\text{old}}$ and, ideally, identical resources and assumptions.
Under this meaning, the threshold of interest, $E_{\text{fixed}}$, is typically an upper bound to the acceptable effort, for instance, the maximum reasonable latency. This case is illustrated in~\autoref{fig:meaning-threshold}, with a threshold defined by the upper bound of the gray area.   

When scalability is the ability to perform better for a range of problem sizes, it corresponds to the function $f_{\text{new}}$ sustaining lower effort than another $f_{\text{old}}$ for all, or most, of the interval of $S$ considered:
\[
    f_{\text{new}}(S; R, A) \leq f_{\text{old}}(S; R, A)
\]
%The function $f_{\text{new}}$ may be better by a factor: $f_{\text{new}}(S; R, A) = c \cdot f_{\text{old}}(S; R, A)$, $c \leq 1$ like the example of \autoref{fig:meaning-better-performance}.
\add{The function $f_{\text{new}}$ may be better by a constant: $f_{\text{new}}(S; R, A) = f_{\text{old}}(S; R, A) - c$, $c > 0$ like the example of \autoref{fig:meaning-better-performance}}.
The key difference with the meaning \textit{threshold} is that there is not necessarily any concern with extending the range of supported problem sizes in that case.

\medskip 

Because the meanings of scalability are connected to the characteristics of $f$, \del{we expect scalability claims in a paper to be supported} \add{the scalability claim in a paper is often supported} by some level of description of the function $f$. We model this aspect by the \textit{level of expression} of $f$, which is closely linked to the methodology employed. The function $f$ may be described by an \textit{explicit function} relating the set of problem size variables $S$ and resources variables $R$ to the set of effort variables $E$. 
%To support claims of scalability depending on the shape of $f$ %=\{e_1, \dots, e_m\}$ such that $e_i = f_i(S, R)$ for $i \in \{1, ..., m \}$.
% \begin{equation}
%     f (S; R, A) = (e_1 = f_1(S, R), \dots, e_m = f_m(S, R)).
% \end{equation}
An explicit function for $f$ may be a model function or an approximation, for instance describing its asymptotic behavior with big $O$ notation. Explicit functions are reported following a complexity analysis, mathematical proofs, or performance modeling. 
The function $f$ may also be described by \textit{sample points} that are measurements of effort variables for a sample of problem size variables. Samples points are reported through examples results, plots, or tables  following a performance evaluation using synthetic data or datasets of varying problem size.

\section{Examples of Model Instantiations}\label{sec:examples}

In this section, we present examples of instantiations of the model using different example papers structured in four stereotypical \textit{scenarios} that are not meant to be comprehensive but rather didactic. The model, including the scalability meaning, corresponds to a scalability claim but scalability experiments/evaluation can also be described in terms of problem size, resources, effort, and assumptions.

\subsection{Algorithm Scalability}
Algorithm papers, including rendering papers, usually present a contribution for which instantiating the model is straightforward. In most cases, the problem size variables are clearly-defined properties of the input data, and the effort variables are fully explained and correspond to the computation cost in terms of execution time or memory. Because algorithm performance can be studied in a controlled manner, it is possible to evaluate their theoretical effort function. Still, most papers use an asymptotic model because the constants vary depending on hardware, configuration, etc.

Consider, for example, the problem of graph drawing that consists of optimizing the layout of the nodes of a node-link diagram on the display. The approach based on solving a stress model requires computing the full matrix of all-pairs shortest paths between the graph nodes. This operation costs $O(|V|^2 \log(|V|)+|V||E|)$ in time and $O(|V|^2)$ in space for a graph with $|E|$ edges and $|V|$ nodes. For this problem, the problem size variables relate to the size of the input graph while the effort variables are the computing time and associated memory consumption. Khoury \etal~\cite{DBLP:journals/cgf/KhouryHKS12} propose an approximation technique to solve this problem in quasilinear time and space and describe its effort function with detailed theoretical expressions of asymptotic complexity, together with experimental measures of computation time for a given implementation. An algorithm is usually considered scalable if it runs in linear time relative to the problem size. 

\begin{cartouche}
    {Drawing Large Graphs by Low-Rank Stress Majorization~\cite{DBLP:journals/cgf/KhouryHKS12}}
    {explicit function (big $O$ notation)}
    {shape (more scalable means more linear)}
    \modelbox
    \modeltwops{number of edges $|E|$}{number of nodes $|V|$}
    \modeltwoef{execution time}{memory consumption}
\end{cartouche}

In some cases, the problem size does not relate to the input data but rather the size of the visual output. 
Falk and Weiskopf~\cite{DBLP:journals/tvcg/FalkW08} consider the problem of rendering 3D vector fields using a texture-based representation. The computational cost of this problem is naturally cubic with the output image resolution. The proposed algorithm uses an image-oriented sampling approach and only computes the parts of the dataset that are represented on the final image. Consequently, this algorithm is mostly independent of the dataset size and predominantly governed by the output resolution and the number of samples. Performance measurements of a GPU implementation show that the rendering times are almost constant as the dataset size increases and scale linearly with the output image and the number of samples.

\subsection{Parallel Computing Scalability}
This scenario is rooted in scaling experiments that are employed to understand the scalability of parallel computing implementations, typically in the context of HPC, cluster computing, or architectures with multiple cores or GPUs. These scenarios also occur in visualization research, mostly in large-data visualization with volume rendering or flow visualization. In this context, scalability is presented as a mapping to compute times with the effort measured for varying problem sizes and hardware resources (\ie compute nodes). 
In this scenario, one major concern is the optimization of computing resource usage, measured using \emph{efficiency}, a metric that compares the gain in execution time (speedup) compared to the amount of additional resources made available to the system. 
The role of varying resources is extensively studied when looking at computational speedup in relationship to number of processors of compute nodes (\eg Gustafson's law \cite{DBLP:journals/cacm/Gustafson88} for weak scalability). This leads to assessing the explicit function of our model, related to resources varying together with problem size.

One such example is by Howison \etal~\cite{DBLP:journals/tvcg/HowisonBC12}. This work assesses the scalability of volume rendering techniques using hybrid parallelism. They measure the problem size in terms of the size of the volume, here, the number of cells in a uniform 3D grid. They discuss that there might be additional data dependency due to the distribution of data values, but identify that there is no relevant impact from data-dependent early ray termination in their case. Effort is measured in memory consumption (MB) and speedup of the compute times (\ie indirectly, the compute times). 
Scalability is primarily understood as the functional mapping to compute times (or speedups). To this end, they measure the effort for varying problem size and hardware resources (\ie compute nodes); in this sense, the functional mapping is represented by quite fine sampling of the function. In addition, they discuss further assumptions, in particular, parameter choices such as various types of block sizes used to distribute the compute work and the type of parallelism (hybrid parallelism vs.\ distributed memory only). 
Moreover, they also evaluate strong and weak scalability.
%Moreover, they also evaluate strong scalability, which is how the compute time varies with the number of compute nodes for a \emph{fixed} total problem size, and weak scalability, which is how the compute time varies with the number of nodes for a fixed problem size \emph{per processor}.

\begin{cartouche}
    {Hybrid Parallelism for Volume Rendering on Large-, \mbox{Multi-,} and Many-Core Systems~\cite{DBLP:journals/tvcg/HowisonBC12}}
    {sample points (fine sampling)}
    {shape (weak and weak-dataset scalability)}
    \modelbox
    \modeltwops{number of cells}{image size}
    \modeloners{number of cores}
    \modeltwoef{compute time}{memory consumption}
    \draw[edge,dashed,<-] (0,-0.5) -- ++(0,-0.4) node[font=\sffamily\scriptsize\itshape] at (0,-1.1) {parameter choices};
\end{cartouche}

% Second example, with graphs
% One of such example is by Tikhonova and Ma~\cite{DBLP:conf/egpgv/TikhonovaM08}. This work assesses the scalability of a parallel graph drawing algorithm that computes the layout in stages. They measure the problem size in terms of input graph size (number of nodes and edges). Effort is measured as the computation time and assessed as visual quality. The authors measure the effort for a set of real-word graphs of various size and complexity. Scalability is understood as the ability to use an increasing numbers of computing resources to process large datasets more efficiently. 

% \begin{cartouche}
%     {A Scalable Parallel Force-Directed Graph Layout Algorithms~\cite{DBLP:conf/egpgv/TikhonovaM08}}
%     {datasets}
%     {shape}
%     \modelbox
%     \modeltwops{number of nodes}{number of edges}
%     \modeloners{number of processors}
%     \modeltwoef{compute time}{visual quality?}
% \end{cartouche}

\subsection{Visual Scalability}
One common scenario of scalability in visualization research targets the study of how the technique's or tool's visual performances are affected by the increase in size of the input data. In this context, scalability is presented as a mapping to readability primarily, and sometimes compute time as well, with the effort measured for varying problem size. These papers present new visualization techniques that can show a larger amount of data in a readable way through data aggregation, interaction, or smart visual encoding. However, they do not always use measurements for the effort function for the readability aspect, but rather discuss the limits of previous encodings, present the rationale behind the new one, and provide visual examples. When the computational aspect of scalability is discussed, it is often in relation to interaction latency.
This scenario seems to be the most typical among visualization papers, at least it is the most commonly found in our survey (see \autoref{sec:literature-analysis}). 
%In this scenario, the computational scalability aspect is discussed in relation to interaction latency.

\begin{cartouche}
    {Structure-aware Fisheye Views for Efficient Large Graph Exploration~\cite{DBLP:journals/tvcg/WangWZSFSCD19}}
    {sample points (example datasets)}
    {better performance}
    \modelbox
    \modeltwops{number of edges}{number of nodes}
    \modeltwoef{computing time}{node overlap}
\end{cartouche} 

Wang \etal~\cite{DBLP:journals/tvcg/WangWZSFSCD19} tackle the visual clutter problem of large graph drawings using a focus+context interactive view. The problem size is the graph size. The main effort variables dependent on problem size are the number of overlapped node pairs which evaluates clutter, and the computing time which evaluates interactivity. The quality of the drawing as evaluated by two metrics (edge orientation offset and shape preservation) and task completion rate are also measured but not in relation to the graph size, \ie as concerns orthogonal to scalability. The authors do not claim that the technique is scalable but rather that their technique is made to mitigate scalability issues and that it outperforms other similar techniques that are compared in an evaluation using five datasets of different sizes. 

\subsection{Cognitive and Perceptual Scalability} 
This scenario covers studies from the psychophysics domain interested in user response to stimuli or studies to investigate the scalability of human perception and cognition when performing a visualization tasks.
In this scenario, studies are close to an ideal controlled environment while being subject to the variability of humans.
The effort variable is the user performance or cognitive load, measured in terms of objectively measurable metrics such as task completion time and accuracy. 
Problem sizes may be input size or task complexity.
In this scenario, articles may present the issues under the terms \emph{perceptual} or \emph{cognitive scalability}, or may not mention \textit{scalable} or \textit{scalability} at all.
In contrast to the visual scalability scenario, in this scenario, studies are not only interested in human limits regarding how visible elements are but also in the cognitive limitations.

Ghoniem \etal~\cite{DBLP:conf/infovis/GhoniemFC04} compare the scalability of two representation techniques for graphs, adjacency matrix and node-link diagram, regarding their readability. The problem size variables are the number of nodes and the edge density, defined as $\sqrt{|E|/[V|^2}$ for a graph with $|E|$ edges and $|V|$ nodes. The effort variable is the readability measured by participants' response time and accuracy across seven tasks such as finding the most connected node or evaluating the edge density. 

\begin{cartouche}
    {A Comparison of the Readability of Graphs Using Node-Link and Matrix-Based Representations~\cite{DBLP:conf/infovis/GhoniemFC04}}
    {sample points (example datasets)}
    {more scalable means faster completion rate on average}
    \modelbox
    \modeltwops{number of nodes $|V|$}{edge density}
    \modeltwoef{task completion time}{accuracy};
\end{cartouche}

Another example by Guiard \etal~\cite{DBLP:conf/bcshci/GuiardBMB01}, investigates the scalability of Fitts' law for pointing tasks, relevant to selection and navigation in any pan-and-zoom environment. Fitts' law describes the empirical relationship between the user movement time and the pointing task difficulty, called index of difficulty $\text{ID}=\log_2(D/W+1)$ where $D$ is the target distance and $W$ the target size. The time for pointing is $\text{MT} = a + b\times \text{ID}$. In the physical world, the ID is limited to about 10 bits. Using a zooming interface in a virtual world, the study shows that Fitts' law still applies beyond the limit.
The effort variable here is the \emph{throughput}, defined as the ratio between the ID and the movement time while the problem size is the ID. 
The study concludes that, for a higher range of IDs, the throughput of multiscale pointing is constant, so Fitts' law holds, limited only by human fatigue.

\subsection{Non-stereotypical and Composite Scenarios}
While these four scenarios are representative of a large part of the typical scalability concerns, we acknowledge that they are not exhaustive: some research work will lay at the intersection of scenarios and others may not exactly fit any scenario.

For systems handling many users, either interacting independently or in direct or indirect collaboration to complete a common task, scalability is understood as the capacity to accommodate for more users. In Glemarec~\etal~\cite{DBLP:conf/vrst/GlemarecBBLLLC19}, the main challenge is to handle multiple user sessions at once: the problem size here is the number of users, while the effort variable is the system performance in terms of latency. In Deng~\etal~\cite{DBLP:conf/chi/DengRKBBF14}, the challenge is to collect a large training set of images with labels indicating the presence or absence of a set of objects of interest, by querying humans on a crowdsourcing platform. Here, the problem size is made of the number of images and the number of objects of interest, which can both be large and users of the system are technically the resources available to complete the annotation task for a dataset. The effort variable is the number of human queries, which directly correlates with the financial cost of completing the annotation task for a dataset. 

Similarly, systems using non-conventional display sizes (mobile displays or tiled displays) sometimes take interest in scalability with the number of composing screens/projectors or the overall display size. The dependent variables describing effort can be related to rendering characteristics like pixel error and compute time (in the form of frame rate) or user task completion time for testing human perceptual or cognitive limits in such configuration (\eg \cite{DBLP:journals/tvcg/YostN06}).

\subsection{Dependent Scalability Issues}\label{sec:dependent}
\begin{figure}
    \centering
    \includegraphics[width=.95\columnwidth]{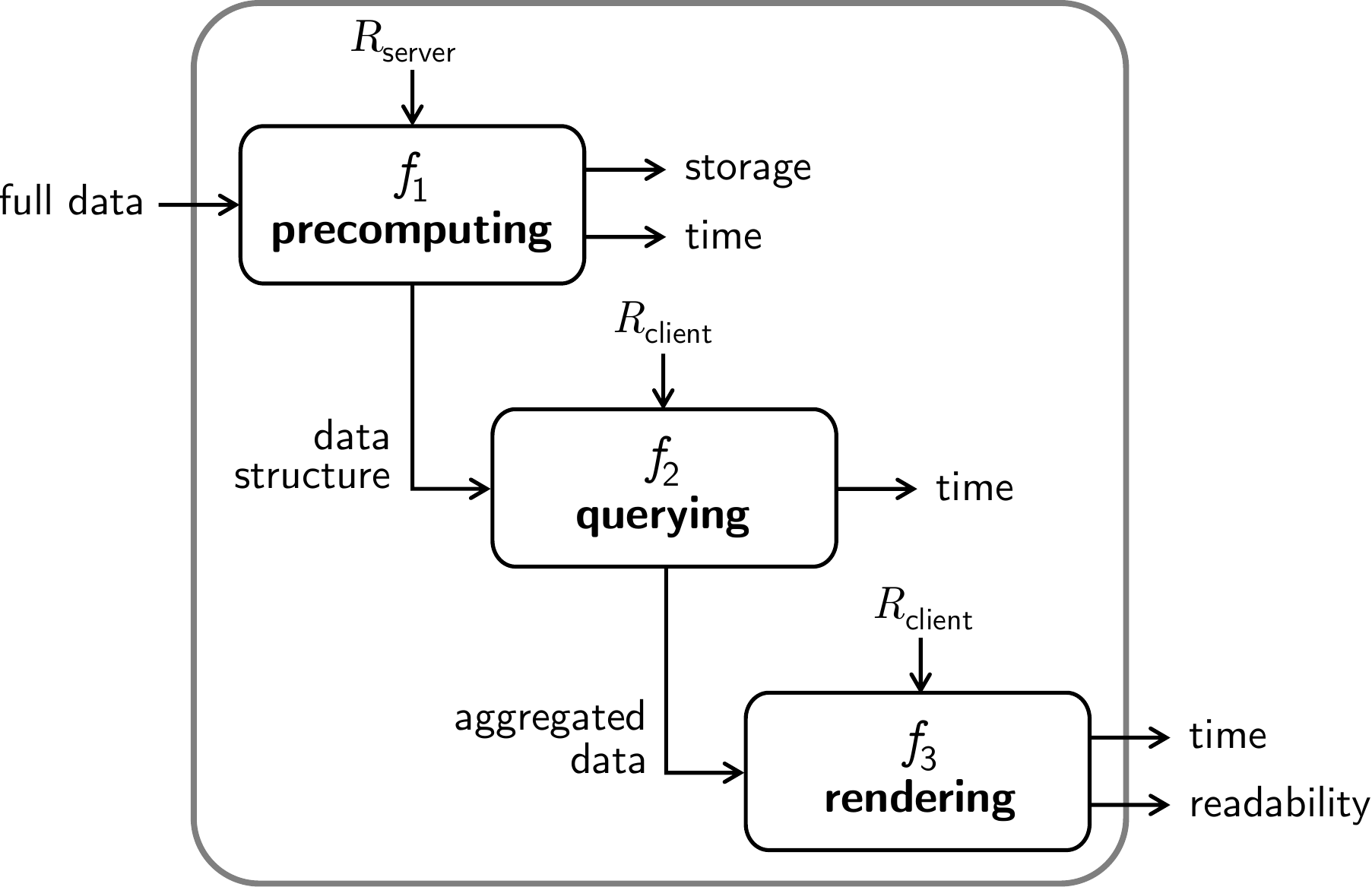}
    \caption{Dependent scalability issues as multiple instances of the model for a common pattern in large, interactive visual analytics systems: aggregate visualization using a precomputed data structure to enable fast interactions. \emph{Rendering} the visualization has no scalability concern since it is ensured, by design, to only handle aggregated data. The focus is on scaling the \emph{querying} step, while the scalability concern for the \emph{precomputing} step varies across applications.}  
    \label{fig:pipeline}
\end{figure}

Visualization system papers may include multiple aspects of scalability (visual, computing) or multiple components (querying component, rendering component) for which the scalability concerns differ and may depend on each other. We focus on this case to show how multiple scalability concerns can be modeled using multiple instances of the model, as illustrated in \autoref{fig:pipeline}.

A common pattern in large, interactive visual analytics systems is to render aggregated data and rely on a precomputation step to build a data structure supporting fast queries for interactive exploration. Some examples among many are data cubes used for multidimensional and spatiotemporal data exploration (\eg Nanocubes~\cite{lins2013nanocubes}), prefetched indexes for brushing and linking interactions~\cite{moritz2019falcon}, or custom in-memory hierarchical structures for detail-on-demand interactions~\cite{DBLP:journals/tvcg/AbelloHK06}.  
In these systems, the scalability concern of the rendering step is trivially solved by representing pre-aggregated data, in the form of heatmaps, histograms, or clustered graphs for instance. Detail-on-demand interactions are based on the querying system that is, in turn, the focus of the scalability concern for increasingly large data. The use of a precomputed data structure to speedup queries trades performance improvement for query time at the cost of the precomputation step and the usage of additional memory to store it. 
Across systems, the assumptions and concerns differ regarding this preliminary step. 
In Nanocubes~\cite{lins2013nanocubes}, the storage size of the data structure is a primary concern since the target is to store it on a laptop. However, their computing time is of lower concern, suggesting data is assumed to be static. 
%In Falcon~\cite{moritz2019falcon}, the indexes are only partial, avoiding the storage concern, and computed dynamically depending on user interactions. Here, query latency is generally mitigated by a prefetching mechanism. However, index computation (the precomputation step) occurs in the client and remains costly for large datasets, even when part of the computation is done on an external scalable database system.
In ASK-Graph~\cite{DBLP:journals/tvcg/AbelloHK06}, no scalability concern is listed for the precomputation step, however, the data structure computation is parameterized by the available resources: available RAM and maximum number of edges that can be processed in a few seconds.

Scalability concerns can also vary as research advances. For instance, for spatial and multidimensional data cubes, HashedCubes~\cite{pahins2017hashedcubes} is presented as an improvement over the state of the art regarding storage size and query time. In addition, they also present building time as an important factor, unlike previous work, and mention supporting dynamic data as future work.

This illustrates that the criteria of scalability can be different across the components of a system (precomputing, querying, rendering), sometimes functioning together. They also depend on the context of application, \eg static vs.\ dynamic data, or client resource requirements. In turn, the amount of space dedicated to these different levels of scalability concerns vary in articles.

\subsection{Comparison with Related Work}

 Related work mentioned factors and measures; they become our problem size, resources, and efforts. The distinction between factors and resources is clarified in our model. The \emph{nuisance variables} can belong to problem size or assumptions, depending on their nature. Regarding multi-criteria optimization advocated by Duboc \etal~\cite{duboc2006scalability}, our efforts could be expressed with a combination of criteria if desired.
In addition, our model also specifies different meanings and expressions of scalability that are usually implicit in each community or domain but can be hard to understand when the readers are from very different backgrounds. Finally, visualization and visual analytics are evolving domains in search of better evaluation methods. Therefore, the meaning and expression of scalability will evolve and require clarification. %This is why we believe our model will help support the evolution of visualization and visual analytics.
We that believe our model can be a first step to encourage the diversification of scalability definitions and help support the development of scalability evaluation methods, especially for research related to the \emph{Visual Scalability} and \emph{Cognitive and Perceptual Scalability} scenarios.

\section{Literature Analysis}\label{sec:literature-analysis}

\newcommand{\arscolor}{green}
\newcommand{\cpscolor}{orange}
\newcommand{\pcscolor}{red}
\newcommand{\vscolor}{blue}
\newcommand{\ars}{ALGO} %{ARS} 
\newcommand{\cps}{PERC} %{CPS}
\newcommand{\pcs}{PARA} %{PCS}
\newcommand{\vs}{VISU} %{VS} 

To relate our model and scenarios to the current state of visualization research, we conducted a structured and systematic literature analysis based on a coding scheme describing the different components of the scalability model.
% We systematically reviewed a sample of the literature from 1990 to 2020. We first formed the sample by automatically filtering papers relevant to the subject using the \texttt{vispubdata} dataset~\cite{Isenberg:2017:VMC}.
We carried out iterative coding rounds to form and refine a \textit{coding scheme} describing the meaning and reasons supporting scalability claims in visualization.
Methodologically, we followed a manual coding process that originally stems from qualitative research and allows for the systematic and structured analysis of literature under ill-defined tasks that necessitate human coders~\cite{charmaz2014constructing}. Such approaches have been frequently used in visualization research, for instance, to characterize evaluation methods~\cite{DBLP:journals/tvcg/LamBIPC12,DBLP:journals/tvcg/0001ICSM13}, interactive dimensionality reduction~\cite{DBLP:journals/tvcg/SachaZSLPWNK17}, and ensemble visualization~\cite{sedlmair2014visual}.
In the following, we describe our methodology, analysis process, and quantitative and qualitative results.

\begin{figure}
    \centering
    \includegraphics[width=.9\columnwidth]{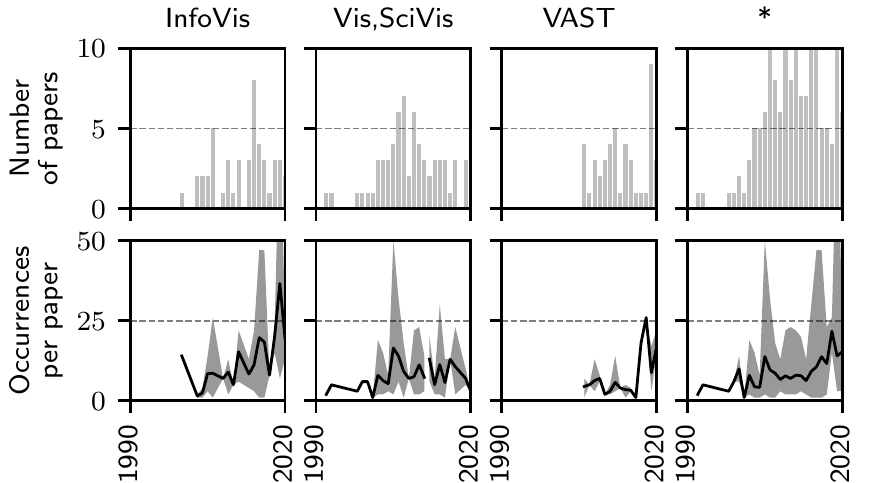}
    \caption{Number of papers and full-text occurrences (band: min-max, line: mean) of the terms \textit{scalability} and \textit{scalable} in the sample of \ncoded~papers, per year. The first three columns show charts per conference, the last column shows charts for the whole. }
    \label{fig:occ-per-year}
\end{figure}

\begin{figure}
    \centering
    \includegraphics[trim=0.25cm 0.25cm 0.25cm 0.25cm,clip,width=\columnwidth]{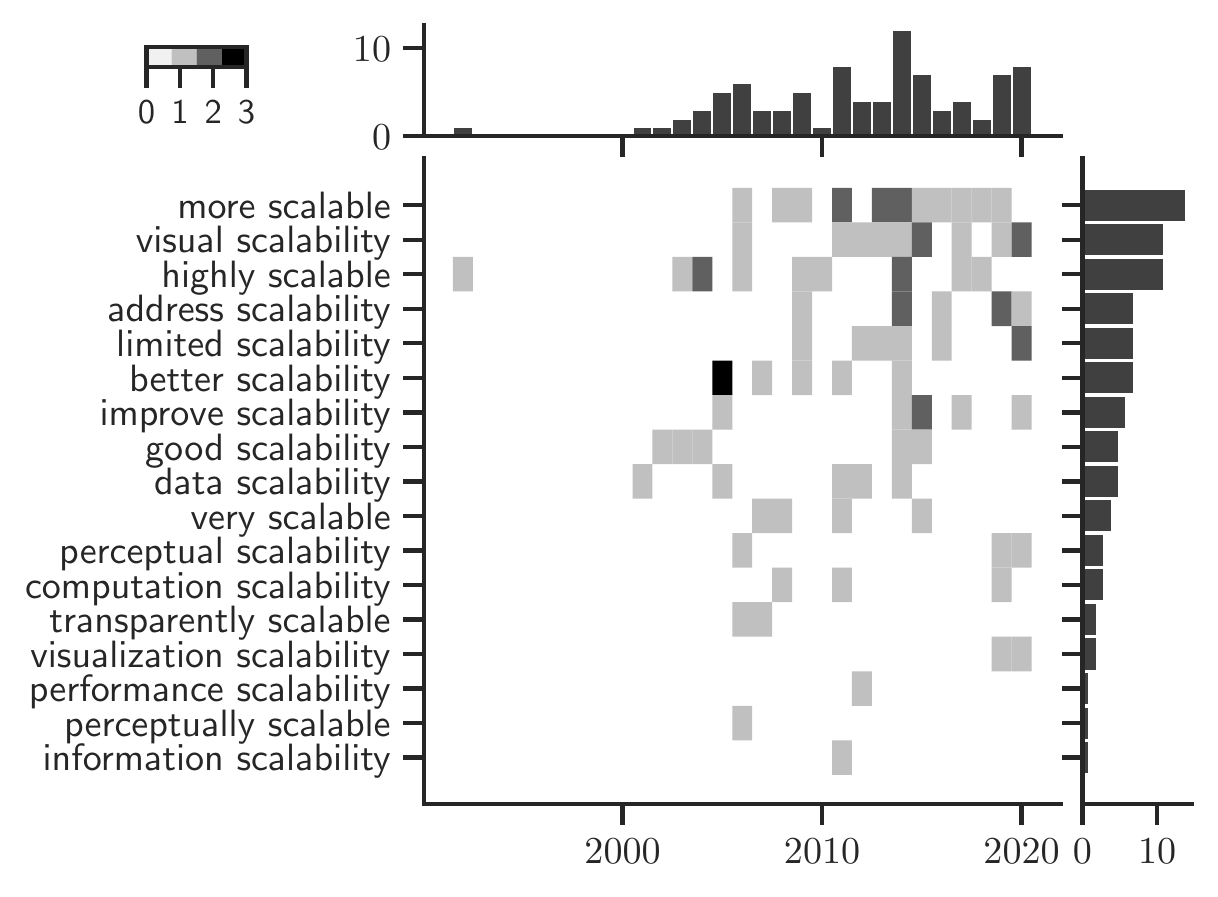}
    \caption{Top-17 occurring expressions including the terms scalability or scalable with number of papers from the sample of \ncoded~papers that use them in their full-text content across years and all papers. }
    \label{fig:top-2-grams}
\end{figure}

\subsection{Literature Sample}\label{sec-lit-sample}
The selection of papers included in our study is based on the \texttt{vispubdata} dataset~\cite{Isenberg:2017:VMC}, containing \totvispub\ papers from the IEEE Visualization (VIS) conferences from 1990 to 2020 (conference and journal articles as well as panel or poster papers). This initial set of papers was filtered automatically to select papers including the prefix \textit{scalab-} at least once in their abstract, title, or keywords in order to capture papers for which scalability was a main concern. This process generated a set of \nfiltered\ relevant papers. After removing \nexcluded\ papers because they mentioned scalability but did not discuss it, we were left with a corpus of \ncoded\ papers: 35 conference, 87 journal, and 5 others articles, among which 6 were of length 5 pages or fewer (\add{full list available at \texttt{\url{https://osf.io/xrvu7/}}}). \autoref{fig:occ-per-year} gives an overview of the frequency of how often the respective terms were used in these papers, organized by year and conference. \autoref{fig:top-2-grams} shows the frequency of the top-17 occurring expressions related to scalability. This list gives a picture of the different aspects and meaning of scalable and scalability. %Out of the  \ncoded\ papers left for the coding process, \nconsensus\ required discussion between coders to reach sufficient agreement. \autoref{tab:summary} gives an overview of the number of collected and coded papers.

\subsection{Coding}

The coding process started with an open-coding phase during which four co-authors of the paper, who developed the model, reviewed 32 seed papers to identify existing variables for the components of our conceptual model for scalability. This phase allowed us to get an overview of the selected literature and led to the development of categories of fixed codes constituting a coding scheme. After multiple iterations, the coding scheme was tested by the four coders on a small sample of 12 papers. The coding scheme is structured by the two objectives of our literature review, which are to describe (1)~\textbf{what} the scalability claim is about in a paper, and (2)~\textbf{how} scalability is presented in the paper. The coding scheme has four categories of codes for the former, and two for the latter:

\begin{description}
    \item[Input] covers the problem sizes and resources of the effort function associated to the paper's scalability claim, \ie they describe what varying parameters are considered, jointly to simplify the coding book. Inputs can be multiple:  \textit{Data Size}, \textit{Data Characteristic}, \textit{Compute Nodes}, \textit{Display Resolution/Units}, \textit{User/User Sessions}.

    \item[Output] covers the outputs of the effort function, \ie the costs or dependent parameters considered in the scalability claim. Outputs can be multiple and may be qualitative (not measured): \textit{Compute Time}, \textit{Memory Consumption}, \textit{User Performance}, \textit{Error/Quality}, \textit{Clutter/Readability}.

    \item[Meaning] covers three different, mutually exclusive, meanings for \textit{being scalable}, defined in relation with our model's effort function. \textit{Function Shape} characterizes scalability based on the shape of the function (constant, linear, bounded). \textit{Extend Domain} defines being scalable as being able to handle problems of larger size than before. \textit{Better Performance} defines scalable as the exhibition of better performance compared to another technique, \ie having a lower effort for the same inputs.

    \item[Scenario] covers the four scenarios from \autoref{sec:examples}: \textit{Algorithm \& Rendering Scalability} (\add{\ars}), 
    \textit{Parallel Computing Scalability} (\add{\pcs}), \textit{Visual Scalability} (\add{\vs}),
    \textit{Cognitive \& Perceptual Scalability} (\add{\cps}). We hypothesize that these scenarios will be able to capture/separate the different definitions of scalability and practices of visualization subcommunities.

    \item[Expression] describes how explicit the effort function is in the paper: \textit{Model Function} and \textit{Asymptotic Function} for function expressions, \textit{Plots/Tables} and \textit{Few Samples} for sample points.

    \item[Reasons] gives possible arguments exposed, or evaluations conducted to support the presented expression of the effort function and associate meaning of the scalability claim: \textit{Unspecified}, \textit{Didactic/Argumentative}, \textit{Theoretical Validation}, \textit{Experimental Validation}, \textit{Case Study/Examples}.
\end{description}

\begin{table}
    \caption{Summary counts for the corpus. Left: Conference counts for the collected and coded papers. Right: Counts for papers excluded, with or without consensus coding, and edge cases. }
    \label{tab:summary}
    \setlength{\tabcolsep}{5pt}
    \begin{tabularx}{.4\columnwidth}{lrr}
        \toprule
        Venue   & Coll.      & Coded    \\ \midrule
        VAST    & 44         & 37       \\
        Vis     & 47         & 36       \\
        SciVis  & 19         & 16       \\
        InfoVis & 47         & 38       \\ \midrule
        Total   & \nfiltered & \ncoded  \\ \bottomrule
    \end{tabularx}~
    \begin{tabularx}{.55\columnwidth}{lrr}
        \toprule
        Set          & Count         & Edge Cases\\ \midrule
        Excluded     & \nexcluded\   & -\\
        Coded        & \ncoded\      & 14\\  
        \hspace{2ex} Consensus    & \nconsensus\  & 5\\
        \hspace{2ex} No consensus & \nredundant\  & 9\\ \midrule
        Total                     & \nfiltered    & \nbarely\ \\
        \bottomrule
    \end{tabularx}
\end{table}

During a second phase, the two other co-authors, who did not participate in the preliminary phase leading to the coding scheme, coded the complete set of papers, to validate the coding scheme and quantitative insights from the literature. These new coders were first introduced to the model and trained on four examples. Some aspects initially part of the coding scheme (number of comparisons with other techniques, assumptions) were removed for simplification and \textit{User/User Session} added to input codes. The coding process progressed in batches, identical for both coders, starting with the 12 papers previously coded for calibration. After each batch, the coders reviewed low-agreement papers to obtain a consensus coding. During that process, they  excluded \nexcluded\ papers for lack of details about scalability and identified \nbarely\ ``edge case'' papers that barely fit into the coding scheme. %These papers were analyzed in a separate open-coding process to learn about the reasons why they did not fit (see \autoref{sec:excluded-papers}).
Out of the \ncoded\ papers from the coded corpus, \nconsensus\ were assigned a consensus coding and \nredundant\ were assigned the average of the two coders' coding (see \autoref{tab:summary}). 
The inter-coder agreement was .72 for the \ncoded\ papers when including the initial low-agreement codes of the \nconsensus\ papers that benefited from a consensus coding, and .76 for the \nredundant\ papers that did not   
%The inter-coder agreement was .72 for all \ncoded\ papers before consensus, and .76 when only considering papers for which no consensus was established 
(Bennett, Alpert and Goldstein's S~\cite{Bennett_S} with Jaccard distance). Using the interpretations of Cohen's kappa, these scores denote substantial agreement (between .61 and .80).

\subsection{Quantitative Results}

\begin{figure}
    \centering
    \includegraphics[trim=0.25cm 0.25cm 0.25cm 0.25cm,clip,width=.95\columnwidth]{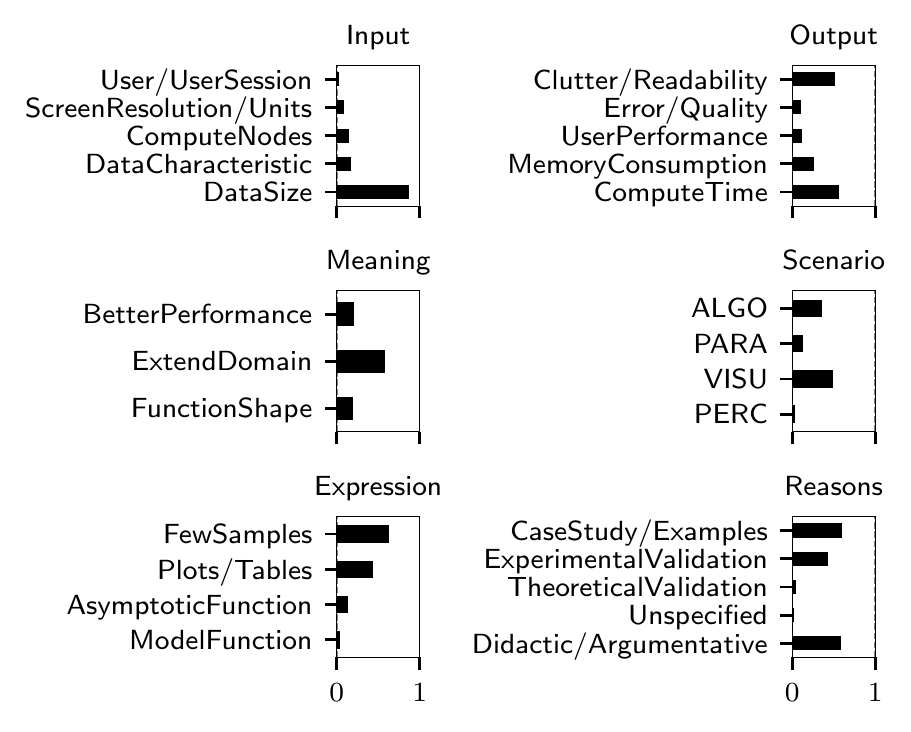}
    \caption{Distribution of codes for the \ncoded~coded papers.}
    \label{fig:distribution-codes}
\end{figure}
% \begin{figure*}
%     \centering
%     \includegraphics[width=\textwidth]{fig/distribution_corpus_codes_h}
%     \caption{Distribution of codes for the \ncoded~coded papers.}
%     \label{fig:distribution-codes}
% \end{figure*}

An overview of the coding results for our corpus is shown in \autoref{fig:distribution-codes}. Most frequently, scalability claims are related to \textit{Data Size} (Input), and concerned with \textit{Compute Times} and/or \textit{Clutter/Readability} (Output). The most typical scalability claim is \textit{Extend Domain} (Meaning). The most frequently represented scenarios are \textit{Algorithm \& Rendering Scalability} and \textit{Visual Scalability} (Scenario). The distribution of the expression codes follows their level of care for strictness with the majority of papers reporting at least some aspect of scalability with singular examples (\textit{Few Samples}), whereas very few described the effort function with the precision of a \textit{Model Function} (Expression). Although linked to the common big $O$ notation for describing algorithm complexity, the \textit{Asymptotic Function} code remains rarely reported. The most commonly found (non-exclusive) reasons to support scalability claims are \textit{Didactic/Argumentative}, \textit{Case Study/Examples}, and \textit{Experimental Validation}.

\begin{figure}
    \centering
    \includegraphics[trim=0.25cm 0.25cm 0.25cm 0.25cm,clip,width=\columnwidth]{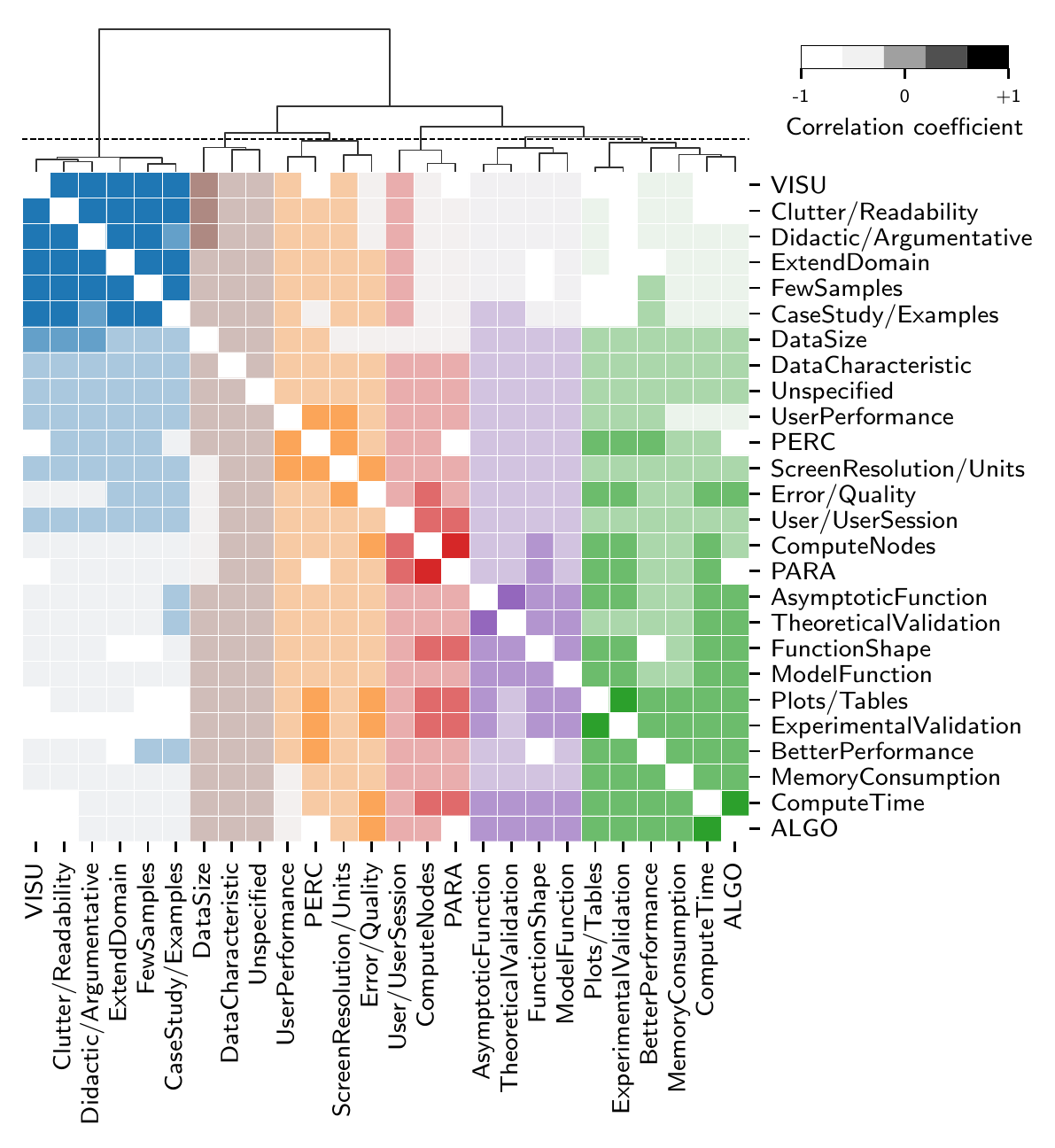}
    \caption{Correlation between codes as a clustered heatmap. Rows and columns are ordered identically, following the order of the dendrogram leaves. Cell hue indicates cluster membership following the six clusters from the represented dendrogram cut (hierarchical clustering: Ward linkage criteria), while cell intensity encodes correlation coefficient between codes (Pearson coefficient).}
    \label{fig:code-correlation}
\end{figure}

\begin{figure}
    \centering   
    \begin{overpic}[trim=0.25cm 0.25cm 0.25cm 0.25cm,clip,width=.95\columnwidth]{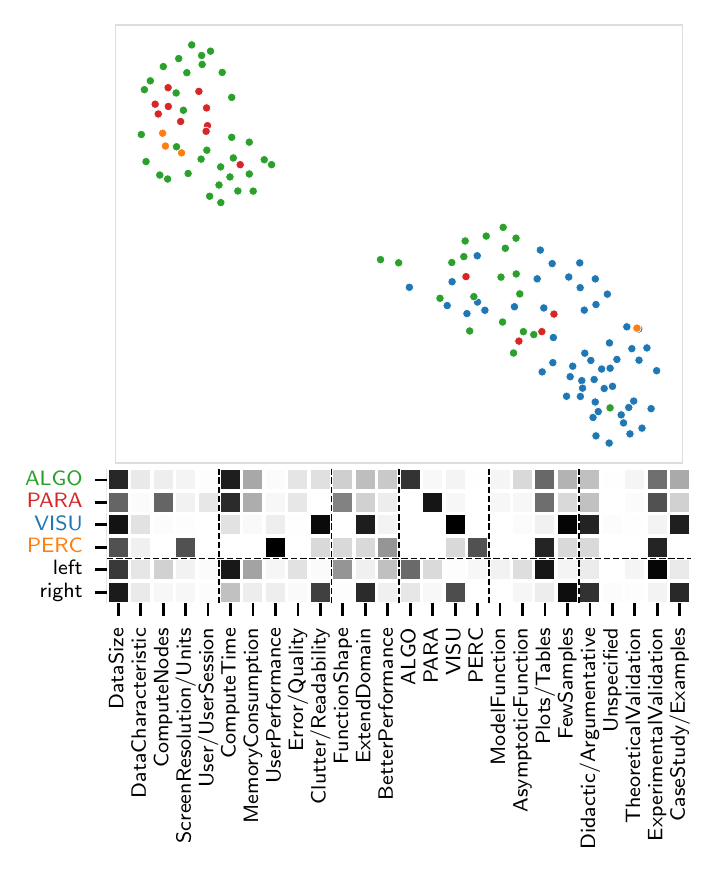}
       \put(0,5){\includegraphics[width=.7cm]{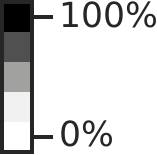}}  
    \end{overpic}
    \caption{Overview of the coded corpus. Top: UMAP embedding of the \ncoded~papers of the corpus based on their codes (excluding the \textit{Scenario}) and colored by scenario. Bottom: distribution of papers codes per coded scenario and per projection cluster (left/right). Cell intensity indicates the proportion of papers from the row group that have the column code: black indicates all, white none.} 
    \label{fig:overview-scenarios}
\end{figure}

To highlight the relationship between the codes of the coding scheme and identify those that are consistently used together, we look at the correlation between codes across the corpus. \autoref{fig:code-correlation} shows the clustered correlation matrix of the codes for the \ncoded~papers. We find the most common coded scenario, \add{\vs}\ (\textit{Visual Scalability}), is linked to the codes \textit{Clutter/Readability}, \textit{Didactic/Argumentative}, \textit{Case Study/Examples}, \textit{Few Samples} (in \vscolor), which describe almost all aspects of the typical visual scalability scenario. The \arscolor~cluster highlights the codes that are related to the \add{\ars}\  scenario, such as \textit{Model Function} and \textit{Function Shape}.
The \cpscolor~cluster, less defined, includes the \add{\cps}\ scenario with some related codes such as \textit{User Performance} (Output). We also find several, unsurprising, pairs of highly correlated codes such as \textit{Compute Nodes} and \add{\pcs}\ (\pcscolor), or \textit{Asymptotic Function} and \textit{Theoretical Validation} (purple).

\autoref{fig:overview-scenarios} presents an overview of the papers. At the top, papers are represented as a UMAP~\cite{UMAP} 2D~projection of their coding and colored by coded scenario. The UMAP projection reveals a clear split between two clusters. At the bottom, the average coding per scenario and per cluster is represented as a heatmap. 
The cluster on the right resembles papers approaching scalability primarily from a visual angle, as characterized in the \add{\vs}\  scenario (\vscolor), with \textit{Clutter/Readability} as an effort variable. The papers from the other scenarios in this cluster are likely those that discuss aspects of scalability that are primarily associated with the \vscolor\ scenario (\ie \textit{Didactic/Argumentative}, \textit{Case Study/Examples}, \textit{Few Samples}). For example, three of the \pcscolor\ papers are related to tiled-displays (screen or projectors). The cluster on the left covers papers approaching scalability primarily from a compute angle, as characterized in the \add{\ars}\ (\arscolor) and \add{\pcs}\ (\pcscolor) scenarios, with \textit{Compute Time} as an effort variable. The papers from the other scenarios in this cluster are those that discuss aspects of scalability that more often associated with the \arscolor\ scenarios (\ie \textit{Plots/Table}, \textit{Experimental Validation}).
Overall, this overview shows that multiple aspects of scalability coexist in the community, and even within our four stereotypical scenarios. It also hints to categorizations of scalability papers finer than, or different from, our scenarios.

\subsection{Results per Scenario} 
The four scenarios were devised after reviewing part of the corpus papers, being aware that some papers will differ from stereotypical cases. In some cases, the scenario in which the scalability claim was demonstrated is different from the scenario of the overall paper. For example, some of the papers that belong to the \add{\ars}\ scenario demonstrated the scalability using parallel implementations (\ie \add{\pcs}\ scenario) or by conducting a user study (\ie \add{\cps}\ scenario). During the coding phase, coders assigned a single scenario per paper and picked the one that matched more closely the overall paper context. While the majority of papers were successfully coded into one of the four stereotypical scenarios, the coders also found recurrent types of papers within each scenario.

In the \add{\ars}\ scenario, the typical paper presented rendering algorithms for scientific data describing biomedical, spatial, or physical phenomena. As shown by their average coding in \autoref{fig:overview-scenarios}, these papers typically have \textit{Data Size} and/or \textit{Compute Nodes} as input, and  \textit{Compute Time} and/or \textit{Memory Consumption} as output. To measure the scalability of the proposed technique or method, they usually conduct an extensive \textit{Experimental Validation} by varying the input parameters and the results is often communicated in forms of \textit{Plots/Tables}. Another type of papers from this scenario address multi-projector displays (6 papers). In these papers, scalability applies to a new calibration method and is relative to an increase in the number of projectors, which challenges the quality of the final picture. Here, the encoded input corresponds to \textit{Screen Resolution/Units}, while the encoded output corresponds to \textit{Error/Quality} and, in some cases, \textit{Compute Time}. In these papers, the scalability claim is often demonstrated by showing few pictures of the final tiled display (\ie \textit{Few Samples}).

The typical paper in the \add{\vs}\ scenario introduced a new visualization technique, or a novel visual analytics tool that aims at solving specific analysis tasks for targeted domain users. As shown by the average coding for \add{\vs}\ papers in \autoref{fig:overview-scenarios}, the typical coded input and output are \textit{Data Size} and \textit{Clutter/Readability}, respectively. The typical scalability claim is demonstrated by a \textit{Case Study} combined with \textit{Didactic/Argumentative} discussion based on few and sometimes only one dataset and has the meaning of \textit{Extended Domain}.  
About half of these papers address the scalability problem by employing summarization, aggregation, or sampling strategies on the data layer, the presentation layer, or both and often in combination with interaction techniques (27 papers).
The rest address the scalability problem with new techniques (\eg layout, interaction) or new visual analytics systems combining existing visualization methods, but without aggregation.

Papers in the \add{\pcs}\ scenario typically describe a new parallel implementation of an existing algorithm, or a new system using a parallel architecture that can scale up to numerous \textit{User Sessions} without introducing undue delay (\ie load scalability). 

Finally, only 5 papers matched the \add{\cps}\ scenario. These papers are generally concerned with measuring \textit{User Performance} relative to varying \textit{Screen Resolution/Units} and \textit{DataSize}, one example being the study of perceptual scalability on large tiled-display walls. While no single scalability meaning emerges, all papers present \textit{Experimental Validation} with results communicated in the form of \textit{Plots/Tables}.   

\subsection{Edge Cases}
\label{sec:excluded-papers}
Throughout the coding process, the coders marked \nbarely~papers as difficult to code using the coding scheme. These papers were additionally open coded for the reasons why they did not fit, to evaluate the limitations of the coding scheme and the conceptual model. After an open-discussion session with all co-authors, we identified four different, non-exclusive, reasons of difficulty:
\begin{enumerate}
    \item \textbf{Require inference:} 7 papers presented a scalability claim (often in the beginning) without establishing a clear link between scalability and the results in the rest of the paper. A common pattern was a switch in the terminology used (\eg using scalability first and then performance).  
    \item \textbf{Use of a scalable related-work component:} 2 papers presented visualization systems relying on a subcomponent or a supporting system said to be scalable. They discussed the scalability of another system, without necessarily relating it to the scalability of their contribution. 
    \item \textbf{Limited scalability:} 1 paper did not claim scalability but rather discussed scalability issues and limitations of their work. While this is a good scientific practice, it was difficult to code, particularly the \textit{Meaning} code.
    \item \textbf{Other meanings:} 2 papers used the word scalability to describe 
    concepts we believe could be best described by other words, \eg adaptability, automation, flexibility. 
\end{enumerate}

\noindent In the latter case, we believe authors could use our model to describe the discussed aspects of scalability and thus, convey a clearer and more meaningful message. While these edge cases did not challenge the coherence of the model, they raised questions about the difference between scalability and other system properties like flexibility.

\subsection{Summary and Discussion}
\label{sec:method:summary}

While the coder agreement confirms the validity and applicability of our model to visualization research, we also observed that the coding scheme fell short of precisely capturing the scalability concern or claim of some papers in our corpus. One reason is that several notions of scalability can coexist in the same paper, for instance connected to different components. For these cases, the paper coding ended up describing multiple aspects as a single effort function, but also grouping different types of reasons together even when they each corresponded to a single effort variable. Although not covered by our coding process, the multiple scalability considerations in a paper could be coded more precisely as different coding instances. Another reason is the lack of consistency in the terminology used that made it difficult in some cases to connect the general scalability discussion to detailed evaluations and results. In other cases, it was not clear if the authors wanted to communicate an improved scalability or scalability limitation through their evaluation.
Our model is meant to address most of these cases, to help authors clarify and expose their claims.

The literature analysis gives an overview of the types of scalability discussed in the corpus and how scalability claims are presented and supported in the corpus of papers. The most frequently represented scenarios are \textit{Visual Scalability} and \textit{Algorithm \& Rendering Scalability} with \textit{Computation Time} and \textit{Clutter/Readability} being the two most common types of effort considered. The meanings of scalable are various even within the same scenario category, the most common being the ability to supporting larger problem sizes than before (\textit{Extend Domain}).

We acknowledge that this overview depends on the balance of topics in the venues chosen for our corpus of papers. Our corpus may be biased toward less of the traditional scalability papers, from the computer graphics community for example, that may be presented at IEEE VIS but originally published in TVCG and not covered by our pool of papers. However, we believe that the IEEE Visualization (VIS) conference publications are, at least for the last decade, representative of the publications in the domain. Moreover, our filtering process also comes with some limitations: similar to our corpus including papers using \textit{scalable} to refer to concepts different to our interest (false positives), some other papers discussing scalability issues under different terms or only in the body of the paper could be missing (false negatives). This could have affected papers from the \textit{Cognitive \& Perceptual} scenario for instance, as they represented only a small portion of our corpus. 
Our filtering process shows that roughly 5\% of the papers were concerned with scalability at the IEEE VIS conferences. To provide some context, we filtered the list of publications from other venues relevant to the visualization domain using the same criteria. We report the numbers and portion of collected publications for these other corpora of papers in~\autoref{tab:other-venues}. Around 5\% or fewer papers are recovered for venues with a broad scope (VIS, EuroVis, TVCG), and around 20\% for the EGPGV and LDAV symposia, which are focused on parallel/large-scale graphics and visualization. This is not surprising since scalability is a major topic in parallel/large-scale visualization, and less so in other conferences with a broader scope. We can anticipate that the proportion of papers discussing scalability in conferences with a broader scope will raise as other visualization subcommunities also develop definitions and methodology to evaluate scalability.

\begin{table}[h]
    \caption{Number of publications collected for different venues, using the same filtering approach as used for our corpus. }
    \label{tab:other-venues}
    \begin{tabularx}{\columnwidth}{Xl}
        \toprule
        Venue                                                    & Collected                              
        \\ \midrule
        IEEE Visualization (VIS) conf. (1990--2020) & 157\textcolor{gray}{/\totvispub} (5\%) \\
        EuroVis (2000--2020)                                     & 44\textcolor{gray}{/1263} (4\%)        \\
        IEEE TVCG (1995--2020)                                   & 209\textcolor{gray}{/3913} (5\%)       \\
        CGF (1982--2020)                                         & 118\textcolor{gray}{/4447} (3\%)       \\
        EGPGV (2002--2020)                                       & 39\textcolor{gray}{/212} (18\%)        \\
        LDAV (2011--2020)                                        & 43\textcolor{gray}{/191} (23\%)        \\
        \bottomrule
    \end{tabularx}
\end{table}

\section{\add{Recommendations and Examples}}
According to our review, the term scalability is used with multiple meanings and the claims of scalability are sometimes difficult to interpret. 
The different visualization subcommunities may have different traditions related to scalability. Most of them borrow methods from other computer science fields such as algorithms and databases. The HPC visualization community is familiar with scalability issues but, when \add{attempting to publish work spanning across subcommunities} \del{publishing at a more diverse venue,} should make sure the well-known HPC/visualization issues \del{they mention} are understood by the others. 
\add{Similarly,} scalability about human issues may be understood by communities connected to HCI or psychology but not always by \eg the computer-graphics community within visualization.
Therefore, we see the need for a clear communication between the wide range of fields and subcommunities that play a role in visualization research.

In the following, we provide a set of recommendations \add{for researchers without a scalability tradition and authors targeting the diverse crowd spanning multiple visualization communities}\del{targeting the diverse crowd of the community, to improve this situation}, together with example papers from our corpus. Clarifying scalability claims will benefit the research process, the readers of the resulting papers, their reviewers, and eventually the visualization research community as a whole. It could even guide future research to cover scalability aspects more thoroughly.

\subsection*{For Researchers} 
We are convinced that considering scalability right from the start, and all the way through, is highly beneficial to a research project.
We envision that doing so is similar to the thorough considerations that researchers already put today into other evaluation questions~\cite{DBLP:journals/tvcg/0001ICSM13,DBLP:journals/tvcg/LamBIPC12,DBLP:journals/tvcg/Munzner09}.

\vspace{2mm}
\begin{description}

\item[Incorporate scalability early on:] Assessing scalability needs planning, and it cannot be done well at a late stage of the work. Choose a scalability goal early and decide how to support your scalability claims. In particular, incorporate scalability considerations in the evaluation of your work. 

\item[Sharpen the expression and meaning of scalability:] Only a few visualization papers try to model or measure a scalability function, \del{and there are only a few that} \add{or} mention the asymptotic behavior. Several papers just report a few measurements. Higher-level characterizations are valuable because they provide a more informative view on scalability. Boosting the description of the scalability expression and meaning has to be included early in the research process.
\end{description}

\subsection*{For Authors}
The following recommendations can help authors improve the presentation of scalability in papers, especially when unfamiliar with scalability issues. Most of these recommendations are a follow-up of the literature review and the difficulty found when parsing the edge case papers. 

\medskip
\begin{description}
\item[Clarify scalability:] Stating or explaining the meaning of scalability in the context of your work can help resolve ambiguity. The explanation could be kept short as long as the meaning becomes clear. This can be done by relating your work to existing and well-established scenarios (\eg~\cite{DBLP:journals/tvcg/JoVDF19}).

\item[Consider our conceptual model:] We expect that many explanations of scalability can be simplified by using our model. In particular, by describing \emph{problem size}, \emph{effort} considered, \emph{resources}, and \emph{assumptions}.

\item[Take similar papers as examples:] Our structured analysis facilitates finding papers that target certain aspects of scalability and, thus, are inspiring presentations for your own work. 

\item[Discuss limits and assumptions:] Many techniques are scalable up to some limit and/or under certain assumptions, and these should be explicitly mentioned.
Similarly, complex systems tend to be hard to evaluate---the evaluation is often restricted to picking just a few measurements. The underlying assumption for the evaluation should also be documented (see ~\cite{Lex_2014,DBLP:journals/tvcg/TiernyGSP09}).
%For example, an algorithm can remain linear in time \emph{provided data has less than 20 dimensions}. A graph layout algorithm can scale linearly \emph{provided the graph is a quasi mesh}.

\item[Do not overload terms:] Do not use the words ``scalability'' or ``scalable'' as synonyms for ``efficient,'' ``fast,'' ``good performance,'' ``faster than baseline,'' etc. Refrain from using them as buzz\-words and use terms consistently to avoid ambiguity. The paper by Abello~\etal~\cite{DBLP:journals/tvcg/AbelloHK06} serves an example where different concepts such as scalability, flexibility, and usability are distinguished.

\item[Match the description and importance of the claim:] Supporting a strong claim of scalability requires explanations, sometimes equations, tables, or figures.  Weaker scalability claims can get by with shorter explanations and less supporting evidence. Choose the right balance according to the importance of scalability in your paper (\eg~\cite{DBLP:journals/tvcg/HadwigerBJP12}).

\item[Consider more than one scalability claim:] A paper may make multiple scalability claims, or present different 
scalability characteristics that are then combined into an overall scalability assessment, as explained in \autoref{sec:dependent}. 
Document scalability for each of the claims, considering the above recommendations (\eg~\cite{DBLP:journals/tvcg/HadwigerBJP12,DBLP:journals/tvcg/BeyerAKLPH13,DBLP:journals/tvcg/IpV11}).

\item[Provide a nil-report:] If a solution to a concrete problem is proposed but does not scale, report it---as an element in a fair assessment of your work. It may encourage others to improve upon your solution (\eg~\cite{DBLP:journals/tvcg/KurzhalsHHBEW16,DBLP:conf/ieeevast/DouWCR11}). 

\end{description}

\subsection*{For Reviewers} 
More explicitly considering scalability can also be beneficial when reviewing papers.

\vspace{2mm}
\begin{description}
\item[Be specific with required revisions:] When asking for more information about scalability, be specific. The above recommendations for paper authors can be used to clarify expectations.
\end{description}

\subsection*{For the Research Community}
Finally, we also see the role of the visualization research community as a whole.
\vspace{2mm}
\begin{description}
    \item[Fostering interdisciplinary communication:] Many communities of visualization, but even more importantly, outside of visualization, already have well-understood interpretations and meanings of scalability. However, these often differ across subfields of computer science. Therefore, an explicit description of the type of scalability implied can help communicate research outside the community. This is particularly important for research that will concern such an outside audience. 
    \item[Include scalability in best practices:] We should strive to improve the way we discuss scalability in our papers, to better compare our approaches and make progress. To this end, existing best practices for conducting and reporting research should be further extended to cover the relevant scalability considerations.
\end{description}

\subsection*{Applying the Model: Coding Examples}\label{sec:how-to}

\begin{table*}[t]
\caption{Coding examples using excerpts from our corpus to illustrate each component of our model.} 
\label{tab:excerpts}   
\begin{tabularx}{\textwidth}{p{6cm}X}
\toprule
\textbf{Input $\,\to\,$ Outputs}  \\ \midrule
% \code{Data characteristics} & \say{we also provide examples to demonstrate that the approach scales well and supports \ul{data of higher dimensions}.}~\cite{DBLP:journals/tvcg/GuoZLLYHMP14}\\
% \code{Data size}, \code{User performance}& \say{We identified three facets of scalability: data size related to the number of data points and categories, perceptual processing related to the ability to perform some tasks efficiently given a data size, and \ul{computation speed} related to the time to compute an image from a visualization technique given a data size.}~\cite{DBLP:journals/tvcg/JoHPKS14}\\
\code{Screen resolution/units$\,\to\,$User performance}&\say{We evaluate the scalability \ul{limits of large, high-resolution, immersive displays} [\dots] Our main metrics concerned \ul{user performance}, specifically elapsed time.}~\cite{DBLP:journals/tvcg/PapadopoulosGK16}\\
\code{Data size, Compute nodes}&\say{[...] is scalable with respect to both \ul{large data sets} as well as \ul{future graphics hardware}.}~\cite{DBLP:journals/tvcg/GeorgiiW06}\\ 
\code{User/User session, Compute nodes$\,\to\,$Compute time}& \say{Our framework should \ul{support many remote user sessions} simultaneously. The performance should \ul{scale under an increased rendering load} as \ul{hardware resources are added}. [\dots] our performance metric is the \ul{turnaround time}.}~\cite{DBLP:journals/tvcg/QiaoMKEK06}\\
% \code{Screen resolution/units}, \code{Compute time}, \code{Error/Quality} & \say{our algorithm scales well to \ul{very large format display walls}, both in terms of alignment \ul{accuracy} and \ul{running time}.}~\cite{DBLP:conf/visualization/ChenSWL02}\\
\code{Screen resolution/units$\,\to\,$Compute time, Error/Quality}&\say{Our approach \ul{efficiently} \ul{scales to projector arrays of arbitrary size} \ul{without sacrificing alignment accuracy}.}~\cite{DBLP:conf/visualization/ChenSWL02}\\
\code{Data size, Data characteristics$\,\to\,$Compute time}&\say{[...] we evaluate our system to assess scalability \ul{in data size} and \ul{data dimension.}}~\cite{DBLP:conf/ieeevast/BrownLBC12}\\
% \code{Data size}, \code{Clutter/Readability}& \say{The proposed method can dramatically increase the scalability of the dendrogram-matrix view. The overview can \ul{easily display over 100 leaf nodes, each taking 10-20 pixels in width}}~\cite{DBLP:journals/tvcg/ChenMP09}\\
% \code{Data characteristics$\,\to\,$Clutter/Readability}&\say{we \ul{can display} \ul{10-20 additional attribute columns} on a desktop display.}~\cite{Nobre_2019}\\
\code{Data size$\,\to\,$Clutter/Readability}&\say{After reordering, adjacent tasks are aggregated to form a single task block, which will significantly \ul{reduce the clutter} in visualization and actually make the schedule visualization more scalable.}~\cite{DBLP:journals/tvcg/JoHPKS14}\\
% \code{Data size}, \code{Clutter/Readability}& \say{On a large desktop screen, we \ul{can show about 50-60 rows}.}~\cite{Nobre_2019}\\
\midrule \textbf{Meaning} \\ \midrule
\code{Better performance}&\say{[...]
we find that Protovis provides \ul{up to 20\texttimes higher frame rates} than prefuse [...]}~\cite{DBLP:journals/tvcg/HeerB10} \\
% \code{Better performance}&\say{we achieve comparable speedups of 2.2\texttimes-3.2\texttimes with 4 GPUs.}~\cite{DBLP:journals/tvcg/JoHPKS14}\\
% \code{Better performance}&\say{As shown in Table 1, Protovis consistently has frame rates an order of magnitude higher, up to 20 times faster for large graph.}~\cite{DBLP:journals/tvcg/HeerB10}\\
% \code{Extend domain} & \say{Our algorithm can render 64 times more data [\dots] at roughly 10 frames per second.}~\cite{DBLP:journals/tvcg/BrivioTC10} \\
\code{Extend domain}&\say{Our study was designed to investigate the perceptual scalability of node-link diagrams for graph connectivity tasks, identifying the graph complexity and \ul{size beyond which they cease to be useful} for such tasks.} \cite{Yoghourdjian_2021}\\
% \code{Extend domain} & \say{The proposed method can dramatically increase the scalability of the dendrogram-matrix view. The overview can easily display over 100 leaf nodes, each taking 10-20 pixels in width.}~\cite{DBLP:journals/tvcg/ChenMP09} \\
% \code{Extend domain} & \say{The detail view can clearly display around 400-450 nodes on an ordinary 1900*1200 screen, with each node taking 4 pixels in width.}~\cite{DBLP:journals/tvcg/ChenMP09} \\
\code{Function shape}&\say{[...] the frame times show \ul{a favorable sublinear scaling} instead of linear scaling as the number of render sessions increase.}~\cite{DBLP:journals/tvcg/QiaoMKEK06} \\
% \code{Function shape} & \say{Notice that both operations scale linearly in data dimensionality.}~\cite{DBLP:conf/ieeevast/BrownLBC12} \\
% \code{Function shape} & \say{Our systematic experiments confirm that for real-life volumetric data this overhead is comparable to the computation of the contour tree demonstrating virtually linear scalability on meshes ranging from 70 thousand to 35 million tetrahedra. Performance numbers show that our algorithm although restricted to volumetric data has an average speedup factor of 6500 over the previous fastest techniques handling larger and more complex datasets}~\cite{DBLP:journals/tvcg/TiernyGSP09} \\
\midrule \textbf{Expression \& Reasons} \\\midrule 
\code{Plots/Table}&\say{As shown in Table 1, Protovis consistently has frame rates an order of magnitude higher, up to 20 times faster for large graphs.}~\cite{DBLP:journals/tvcg/HeerB10} \\
\code{Asymptotic function}&\say{Overall bound: $O(n\log(n) + N \alpha (N) + g \times N S ).$}~\cite{DBLP:journals/tvcg/TiernyGSP09} \\
\code{Case study}&\say{We demonstrate a case study with 276 samples which is considered a large study of mRNA-seq data by current standards [...]}~\cite{DBLP:journals/tvcg/StrobeltABPBPL16}\\
\code{Experimental validation}&\say{[...] we describe additional experiments to evaluate the scalability of Dis-Function in a controlled manner. Specifically, we examine the performance of Dis-Function as the dataset grows in size (in terms of number of rows) and in complexity (in number of dimensions) independently.}~\cite{DBLP:conf/ieeevast/BrownLBC12} \\
\code{Asymptotic function}, \code{Theoretical validation}& \say{These fields can be computed in linear time on the GPU and queried in constant time. Therefore, the complexity of the algorithm is reduced from quadratic to linear.}~\cite{DBLP:journals/tvcg/PezzottiTMHLLEV20}\\
% \midrule \textbf{Assumptions} \\ \midrule
% Bottleneck&\say{The reason behind the non-optimal scalability for larger process counts is because each block is assigned to only one process. Since some blocks will take more time than others, the wall clock time is bounded by the slowest processes.}~\cite{DBLP:journals/tvcg/NouanesengsyLS11} \\
% Loss of details&\say{Such large-scale overview comes at the expense of showing finer details: our design considers how visual principles suggest layouts that allow large-scale patterns to emerge, and includes abstraction mechanisms designed to retain salient features.}~\cite{DBLP:journals/tvcg/AlbersDG11}\\
\bottomrule
\end{tabularx}
\end{table*}

In \autoref{tab:excerpts}, we provide examples of how we coded the paper corpus. We do so by quoting selected excerpts from the selected papers and show how they map to our coding scheme. 
These excerpts are good examples of ways to mention the main model components in a research paper, which we recommend authors to do to present definitions accessible for all visualization research domains.

\section{Challenges and Future Directions}\label{sec:discussion}
Clarifying scalability in visualization goes beyond improving the current state of communicating research results, it also highlights open issues for the visualization research community.

\subsection{Scalability and Evaluation}

Characterizing scalability according to our model is tightly linked to the more general problem of evaluating visualization. There is an ongoing discussion in the visualization community about appropriate ways to perform evaluation~\cite{DBLP:journals/tvcg/0001ICSM13,DBLP:journals/tvcg/LamBIPC12}, even a full workshop dedicated to the topic since 2006  (\href{https://beliv-workshop.github.io/}{BELIV}).

Evaluation is especially hard when we want to assess human-related aspects. This directly relates to the problem of measuring the effort in our conceptual model. With experiments involving human participants, one can arrive at samples of the measured outputs.  However, some performance measurements are notoriously hard to characterize or measure, such as readability and understanding. 

Therefore, it is interesting to use existing or develop new proxies for measurements such as memorability, readability, or discriminability~\cite{Veras2019}. Even harder is the development of computational models for evaluation, or tools to facilitate the evaluation of techniques and systems such as EvalBench~\cite{aigner2013evalbench} and Touchstone~\cite{Touchstone2}. They would be extremely useful to support faster evaluation and enrich the scalability expression, but they can be very difficult to do in general; they apply to visualization or interaction techniques but not to systems. 
However, fundamental evaluation issues are not restricted to human-oriented studies. For example, it is already hard to characterize all relevant input parameters and assumptions for a complex technical system such as a multi-GPU cluster running an advanced volume rendering system. Control of input parameters is also related to the issue of external validity of experiments, for example, for real-world data (uncontrolled) vs.\ synthetic data (controlled, but not realistic). One promising solution could be the extension of the approach of generative data models~\cite{DBLP:conf/beliv/SchulzNEFHHKNSB16}.
This is also linked to developing additional quantification methods for visualization~\cite{Schreiber:2022:QVC}.
In summary, we see the need for further developments in evaluation methods to improve the characterization of scalability. 

\subsection{Reproducibility, Comparison, and Benchmarking}

Addressing scalability in a visualization article is important not only for the paper but also for the community, to build knowledge about the comparative scalability of related research work. We are interested in monitoring the progress in research on scalability over the years, but we still lack consistency in scalability reports and meanings.
Reproducibility and replicability~\cite{ReproVis} help tremendously for these comparisons and the community should incentivize visualization research to increase the number of reproducible articles.  Otherwise, similar articles may decide to report about their scalability with very different choices, defeating or lowering the purpose of reporting scalability. We do not argue for a total alignment of research methods but rather to a convergence toward a consensual set of effort measures, methods to collect them, and to report them, with flexibility in reporting extra measures or using alternative, hopefully, improved methods.

With sufficient maturity, these comparisons should become benchmarks. Proper benchmarking and comparison between visualization contributions are important to drive research in our community~\cite{Benchmarks, battle:hal-02556400}. While these are relevant in general, they lead to specific challenges for scalability research because they need to include variations in problem size and properly control the assumptions. There are a number of (implicit) benchmarks in our community, for example, including the Contests at IEEE InfoVis, VAST, SciVis, and the Graph Drawing conference, but these usually rarely consider scaling the problem size. A related problem is the comparison of the resulting efforts. While this is doable for individual measures (for example, comparing overall compute times), it might be harder for cases that come with multiple measurements (for example, for different components of a complex system; or when assessing several facets such as speed, accuracy, readability, etc.). 
This leads to considering multiple efforts or multi-criteria efforts, mentioned by Duboc \etal~\cite{duboc2006scalability} but new in visualization.

\subsection{Role of Scalability in Future Research}

A quite fundamental question is the extent to which we---as a community---should consider scalability in future research. Should a majority of papers address scalability as a relevant aspect to characterize visualization contributions? Or is it more appropriate to restrict the discussion of scalability to research in core subfields?

The SciVis community has been involved for many years with HPC where scalability is central. In-situ visualization~\cite{In-Situ} is strongly concerned with scalability but in the sense of not interfering with HPC code while providing useful visualizations and sometimes steering.
The VAST and InfoVis communities have no identified subfield addressing scalability explicitly, except progressive visual analytics~\cite{Dagstuhl18411}. 

The community should keep track of its advancements in scalability and report its progress in a more structured way. Exposing progress in scalability would be an incentive and a useful asset for our community. 

\subsection{New Scenarios}

With our quite flexible conceptual model, we want to keep scalability open for further interpretation and scenarios. In particular, we want to specifically avoid boxing-in any future research; that is why we did not aim to define scalability but instead, we provide a generic framework. We envision that new scenarios, or even a new set of scenarios, and best practices will be developed by our research community. This could even include new strategies to improve scalability, for example, new paradigms in computing, alternative ways to deal with various trade-offs, or novel representations. 

One challenging problem relates to summarization techniques in visualization, such as Scagnostics~\cite{Scagnostics} or ``Accordion Drawing''~\cite{PRISAD}, designed to remain readable by selectively showing specific aspects of the data. This approach reduces the problem size drastically, using clustering, sampling, or aggregation techniques, to address the display resolution bottleneck for instance.
We think this is a specific case of scalability claim, as it corresponds to a trade-off between the level of information shown and the problem size. 
More work is needed to  properly characterize the scalability of these multi-aspect techniques; it will require improving how assumptions are reported in addition to identifying the meaningful measures.

\section{Conclusion}

We presented a conceptual input--output model that allows us to characterize different scenarios of scalability in visualization research. 
We used the model as a lens to systematically analyze existing research on scalability in visualization, \add{derive recommendations for communicating scalability across different subcommunities,} and highlight the open issues for scalability in the community. 
We hope that our work will help others, especially in the information visualization and HCI communities, to more easily and precisely characterize their scalability claims, and also to inspire them to conduct more research into scalability definition and associated evaluation methods. 
After all, the increase of data will most likely continue and we, as a community, will need to keep pace by ensuring that our contributions ``scale'' along. %In the future, we hope that \del{the standardization of instantiations of our model will encourage the development of benchmarks to compare scalability in} visualizations research and foster reproducibility.

% use section* for acknowledgment
\ifCLASSOPTIONcompsoc
  % The Computer Society usually uses the plural form
  \section*{Acknowledgments}
\else
  % regular IEEE prefers the singular form
  \section*{Acknowledgment}
\fi

The authors would like to thank the reviewers of our initial submission to IEEE VIS 2021 and the reviewers of this submission for their feedback and suggestions, which helped us improve this article. MA acknowledges support by Deutsche Forschungsgemeinschaft (DFG, German Research Foundation) under Germany's Excellence Strategy -- EXC 2120/1 -- 390831618, and MS and DW acknowledge support by DFG -- Project-ID 251654672 -- TRR 161.
This research was also supported by DATAIA convergence Institute as part of the ``Programme d'Investissement d'Avenir'', (ANR-17-CONV-0003) operated by Inria and IP-Paris.

% Can use something like this to put references on a page
% by themselves when using endfloat and the captionsoff option.
\ifCLASSOPTIONcaptionsoff
  \newpage
\fi

% trigger a \newpage just before the given reference
% number - used to balance the columns on the last page
% adjust value as needed - may need to be readjusted if
% the document is modified later
\IEEEtriggeratref{78}
% The "triggered" command can be changed if desired:
\IEEEtriggercmd{\enlargethispage{-1in}}

% references section

% can use a bibliography generated by BibTeX as a .bbl file
% BibTeX documentation can be easily obtained at:
% http://mirror.ctan.org/biblio/bibtex/contrib/doc/
% The IEEEtran BibTeX style support page is at:
% http://www.michaelshell.org/tex/ieeetran/bibtex/
\bibliographystyle{IEEEtranDOI}
% argument is your BibTeX string definitions and bibliography database(s)

\bibliography{references,corpus}
%\bibliography{references}

% biography section
% 
% If you have an EPS/PDF photo (graphicx package needed) extra braces are
% needed around the contents of the optional argument to biography to prevent
% the LaTeX parser from getting confused when it sees the complicated
% \includegraphics command within an optional argument. (You could create
% your own custom macro containing the \includegraphics command to make things
% simpler here.)
%\begin{IEEEbiography}[{\includegraphics[width=1in,height=1.25in,clip,keepaspectratio]{mshell}}]{Michael Shell}
% or if you just want to reserve a space for a photo:

\begin{IEEEbiography}
[{\includegraphics[width=1in,clip,keepaspectratio]{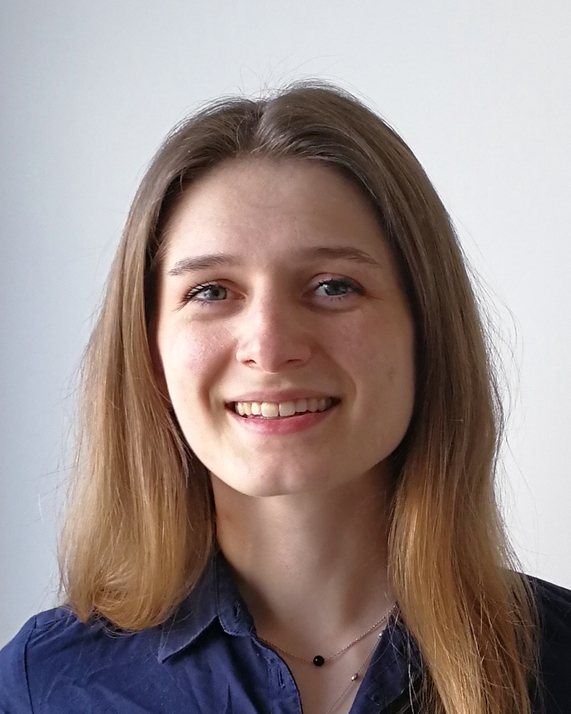}}]
{Ga\"elle Richer}
is a postdoctoral researcher at Inria Saclay, Aviz research team. She received her PhD in Computer Science in 2019 from the University of Bordeaux, France where she worked at the LaBRI. Her research interests include Interactive Visualization and Visual Analytics with a focus on scalability issues.
\end{IEEEbiography}
\vskip -2\baselineskip plus -1fil 
\begin{IEEEbiography}
[{\includegraphics[width=1in,clip,keepaspectratio]{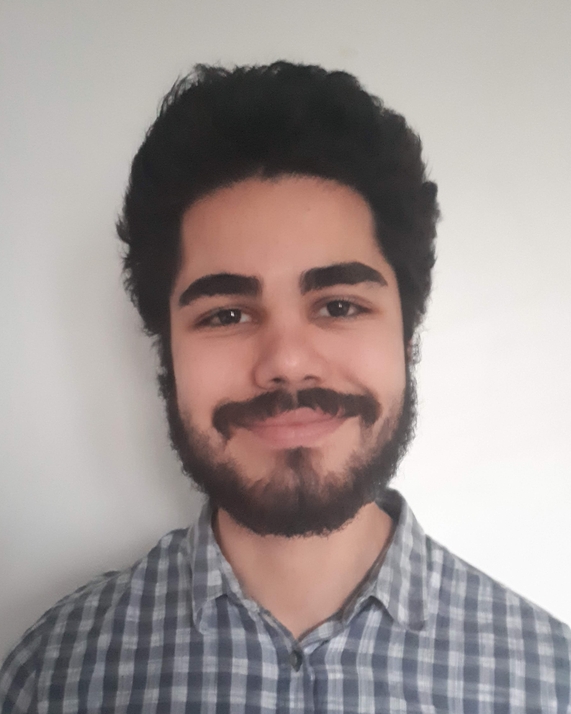}}]
{Alexis Pister}
is a second year Ph.D. student at Inria Saclay Aviz team and Telecom Paris. His subject consists in developing visual analytics tools for the use of historians and sociologists for Social Network Analysis with an emphasis on interpretability and user experience. His research interests are Visual Analytics and Network Analysis.
\end{IEEEbiography}
\vskip -2\baselineskip plus -1fil
\begin{IEEEbiography}
[{\includegraphics[width=1in,height=1.25in,clip,keepaspectratio]{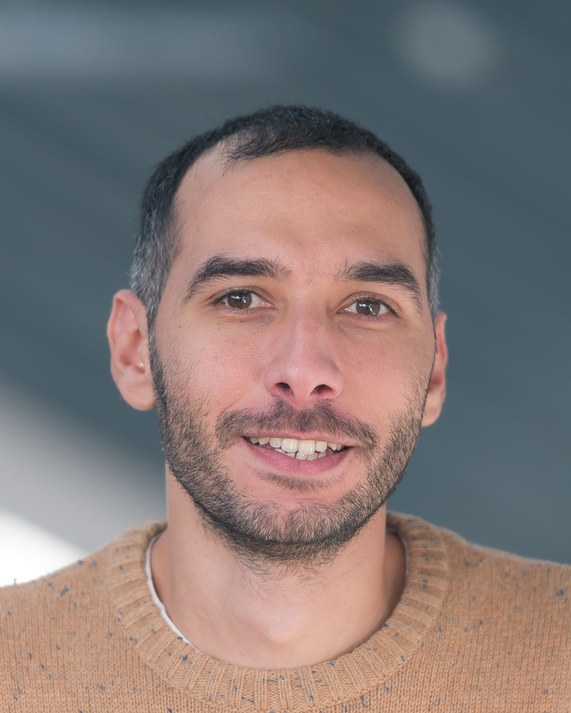}} \raisebox{1ex}{\tiny Photo: Max Kovalenko}]{Moataz Abdelaal}
 is a fourth year Ph.D. student at the Visualization Research Center (VISUS) at University of Stuttgart. His research interests are Network Visualization, Visual Analytics, and Human-Computer Interaction.
\end{IEEEbiography}
\vskip -2\baselineskip plus -1fil 
\begin{IEEEbiography}
[{\includegraphics[width=1in,height=1.25in,clip,keepaspectratio]{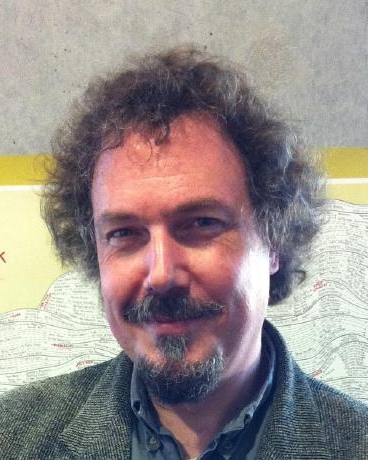}}]
{Jean-Daniel Fekete}
is the Scientific Leader of the INRIA Project Team Aviz that he founded in 2007. He received his PhD in Computer Science in 1996 from University of Paris-Sud, France. His main research areas are Visual Analytics, Information Visualization and Human Computer Interaction. He is a Senior Member of IEEE.
\end{IEEEbiography}
\vskip -2\baselineskip plus -1fil 

\begin{IEEEbiography}[{\includegraphics[width=1in,height=1.25in,clip,keepaspectratio]{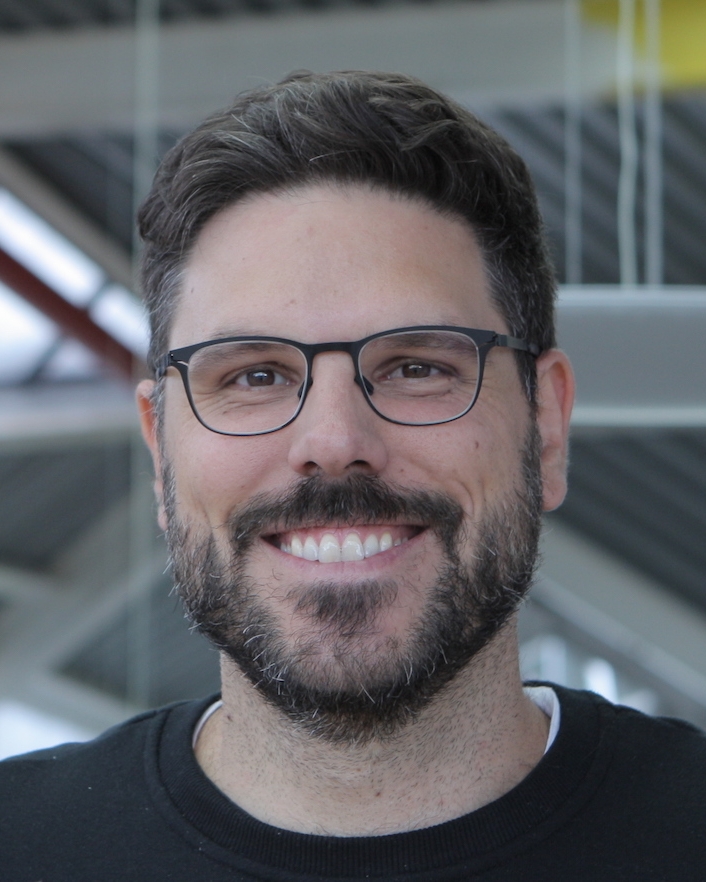}}]{Michael Sedlmair} 
 is a professor at the University of Stuttgart and heads the VisVAR research group there. He received his Ph.D. degree in Computer Science in 2010
from the Ludwig Maximilians University of Munich, Germany. His research interests focus on Visualization/Visual Analytics, Human-Computer Interaction, and Augmented/Virtual Reality.
\end{IEEEbiography}
\vskip -2\baselineskip plus -1fil 
\begin{IEEEbiography}
[{\includegraphics[width=1in,height=1.25in,clip,keepaspectratio]{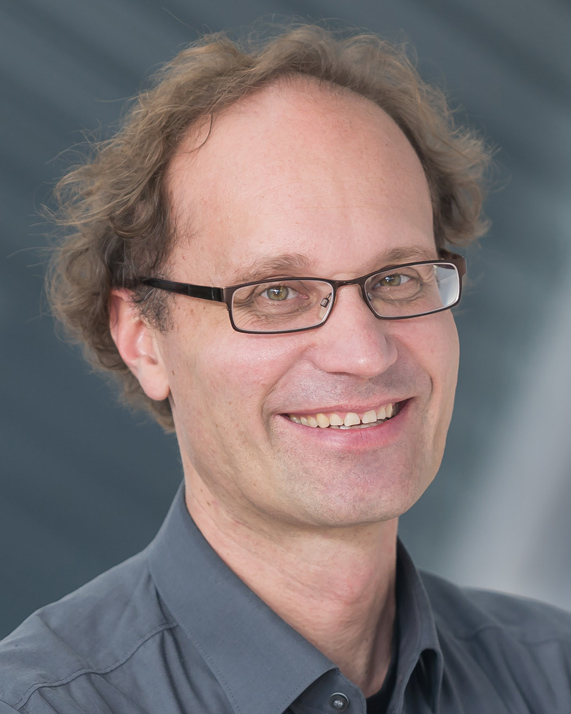}} \raisebox{1ex}{\tiny Photo: Max Kovalenko}]{Daniel Weiskopf}
is a Professor at the Visualization Research
Center (VISUS) of the University of Stuttgart,
Germany. He received his Dr.\,rer.\,nat.\ degree (similar to PhD) in
Physics from the University of T\"ubingen, Germany, in 2001, and the
Habilitation degree in Computer Science from the University of
Stuttgart in 2005.
His research interests include Visualization, Eye Tracking, Human-Computer Interaction, Computer Graphics, and Special and General Relativity. 
%He is a member of the IEEE Computer Society, ACM SIGGRAPH, Eurographics, and the Gesellschaft f\"ur Informatik.
\end{IEEEbiography}
\vfill

% You can push biographies down or up by placing
% a \vfill before or after them. The appropriate
% use of \vfill depends on what kind of text is
% on the last page and whether or not the columns
% are being equalized.

%\vfill

% Can be used to pull up biographies so that the bottom of the last one
% is flush with the other column.
%\enlargethispage{-5in}

% that's all folks
\end{document}